\documentclass{article}
\usepackage{ijcai24}
\usepackage{times}
\usepackage{soul}
\usepackage{url}
\usepackage[hidelinks]{hyperref}
\usepackage[utf8]{inputenc}
\usepackage[small]{caption}
\usepackage{graphicx}
\usepackage{amsmath}
\usepackage{amsthm}
\usepackage{booktabs}
\usepackage{algorithm}
\usepackage{algorithmic}
\usepackage[switch]{lineno}
\usepackage{bbm}
\usepackage{bm}
\usepackage{natbib}
\usepackage{amssymb}
\usepackage{xcolor}
\usepackage{mathtools}
\usepackage{tikz}
\usepackage{setspace}
\usetikzlibrary{3d}
\usepackage{comment}
\usepackage{appendix}

\newtheorem{theorem}{Theorem}
\newtheorem{lemma}{Lemma}

\newtheorem{corollary}{Corollary}
\newtheorem{proposition}{Proposition}
\usepackage{enumitem}
\usepackage{subcaption}
\usepackage{pgfplots}
\pgfplotsset{compat=1.18}
\usepackage[T1]{fontenc}
\title{Simultaneous All-Pay Auctions with Budget Constraints}

\author{
Yan Liu$^1$
\and
Ying Qin$^1$
\and
Zihe Wang$^1$
\affiliations
$^1$Renmin University of China \\
\emails
liuyan5816@ruc.edu.cn,
2021201494@ruc.edu.cn,
wang.zihe@ruc.edu.cn
}

\begin{document}
\maketitle

\begin{abstract}
The all-pay auction, a classic competitive model, is widely applied in scenarios such as political elections, sports competitions, and research and development, where all participants pay their bids regardless of winning or losing.
However, in the traditional all-pay auction, players have no budget constraints, whereas in real-world scenarios, players typically face budget constraints.
This paper studies the Nash equilibrium of two players with budget constraints across multiple heterogeneous items in a complete-information framework.
The main contributions are as follows: (1) a comprehensive characterization of the Nash equilibrium in single-item auctions with asymmetric budgets and valuations; (2) the construction of a joint distribution Nash equilibrium for the two-item scenario; and (3) the construction of a joint distribution Nash equilibrium for the three-item scenario.
Unlike the unconstrained all-pay auction, which always has a Nash equilibrium, a Nash equilibrium may not exist when players have budget constraints.
Our findings highlight the intricate effects of budget constraints on bidding strategies, providing new perspectives and methodologies for theoretical analysis and practical applications of all-pay auctions.
\end{abstract}

\section{Introduction}
Competition is ubiquitous in life.
For example, in political elections, two presidential candidates compete across multiple states, allocating their limited resources to different states in the hope of securing more votes. 
In the development of large language models, IT companies operate under budget constraints and must allocate their limited resources among computing power, data acquisition, and talent recruitment to develop the most advanced models. 
A common characteristic of these competitive scenarios is that ultimately, only one player wins and receives the reward, while every participant incurs costs, which may include money, time, or effort. 
The all-pay auction, as a classic competition model, effectively captures the nature of such competitive situations \citep{nalebuff-1983,hillman-1989}.

In an all-pay auction, all players compete against each other to obtain an item or prize. 
Specifically, each player submits a bid, and the player with the highest bid wins and receives the item. 
The key distinction from other auction formats, such as first-price or second-price auctions, is that in an all-pay auction, each player must pay their bid regardless of whether they win or lose.
Because the all-pay auction provides a compelling model for simulating competitive scenarios, it has attracted significant interest from researchers in fields such as computer science, sociology, and economics \citep{barut-1998,che-1996,dziubinski-2023,avni-2020,avni-2021,dechenaux-2015}.
In the classic all-pay auction, players do not have budget constraints.
However, in real-world scenarios, players are often subject to budget constraints.
Therefore, we investigate the all-pay auction with budget constraints under complete information.
Specifically, the budgets of the two players are asymmetric, and the items are heterogeneous.
Moreover, the value of each item is also asymmetric between the players.
This general setting allows us to more realistically describe the competitive relationships between players.

In the study of all-pay auctions with budget constraints, Roberson's work is the most relevant to ours.
However, in their setting, the items are homogeneous \citep{roberson-2012}.
Another closely related work is that of Dulleck, but in their setting, the resources invested by players only satisfy the budget constraint in expectation, and players aim to maximize their winning probability rather than their utility \citep{dulleck-2006}.
In the above two studies, their conclusions only provide the strategy for players on each individual item but do not offer the joint strategy for players across multiple items.
In our work, the budget constraint is a hard constraint rather than a soft constraint in expectation.  
Moreover, the Nash equilibrium we provide will be the joint distribution across multiple items, rather than merely the marginal distribution on a single item.

However, analyzing the Nash equilibrium in all-pay auctions with multiple items and budget constraints is a highly challenging problem, primarily due to the following reasons: when players do not have budget constraints, they can bid freely on each item, making their bidding strategies more independent.
In contrast, when players face budget constraints, their bids become interdependent, introducing a correlation between their bidding decisions across items.
In summary, when players lack budget constraints, their bids on each item follow an independent distribution, meaning that bids on different items do not directly influence each other \citep{barut-1998}.
However, once players have budget constraints, their bid distributions typically form joint distributions, where bids on different items may be correlated \citep{roberson-2012}.

\subsection{Our contribution}
In this paper, we are the first to study the joint distributions of two budget-constrained players bidding on multiple items with heterogeneous values.
Our contributions are as follows:
\begin{itemize}
    \item For a single item, under the conditions of budget asymmetry and asymmetric valuations, we fully characterize the Nash equilibrium.
    \item We extend our analysis from a single item to two items and construct a Nash equilibrium, which takes the form of a two-dimensional joint distribution.
    \item For three items, under the assumption that item values are symmetric between players, we construct a Nash equilibrium, which is a three-dimensional joint distribution.
\end{itemize}

\subsection{Related work}
There is a rich body of literature on all-pay auctions.
We provide an overview of all-pay auctions with budget constraints under complete information.

\citet{dulleck-2006} consider the Nash equilibrium strategies of two players across multiple items.
However, they only require that the amount of resources allocated does not exceed the budget in expectation, and each player focuses solely on maximizing their probability of winning rather than their utility.
\citet{dekel-2007} analyze how allowing two players to alternate bids, with the possibility of jumping bids, affects auction outcomes.
\citet{roberson-2012} provide the marginal distributions of the Nash equilibrium strategies for two players with asymmetric budgets across multiple homogeneous items.
\citet{hwang-2023} analyze the equilibrium allocation strategies of two players with symmetric budget constraints in continuous items, which approximate an environment with an arbitrarily large (but finite) number of items.

Another competitive model related to the all-pay auction is the Colonel Blotto game \citep{roberson-2006}.
In this model, players also have budget constraints and place bids across multiple items.
For each item, the player with the highest bid wins and obtains the item, but they do not have to pay their respective bids.
As a result, players tend to exhaust their budgets in competition.  
In contrast, in the all-pay auction, since players must pay their bids regardless of winning or losing, they may not necessarily spend their entire budgets.
Researchers have conducted extensive studies on the Nash equilibrium of the Colonel Blotto game and its variants \citep{kovenock-2021,boix-2020}.
However, under the conditions of multiple heterogeneous items and asymmetric budgets among players, solving for the Nash equilibrium remains an open problem \citep{roberson-2006}.

\section{Preliminaries}
We consider a model with two players and $n$ items.
Each player $i \in \{1, 2\}$ has a budget constraint $B_i \geq 0$ and values item $j \in \{1, 2, \dots, n \}$ at $v_{ij} > 0$.
Player $i$'s pure strategy is an $n$-dimensional bid vector $\bm{x}_i = (x_{i1}, x_{i2}, \dots, x_{in})$, where $x_{ij} \geq 0$ represents their bid on item $j$, subject to the budget constraint $\sum_{j = 1}^n x_{ij} \leq B_i$. Denote $\bm{X}_i$ as player $i$'s set of pure strategies:
\begin{align*}
    \bm{X}_i = \Big\{(x_{ij})_{j=1}^n: \sum_{j=1}^n x_{ij} \leq B_i, \text{ and } x_{ij} \geq 0 \Big\}.
\end{align*}

We consider a setting where players have full information about each other's values and budgets and choose their strategies simultaneously\footnote{This full-information setting serves as a baseline for analysis before introducing uncertainty or incomplete information. It is also practical in certain industries due to regulatory transparency, structural disclosure requirements, and the availability of historical data and valuation frameworks.}. In the game, the player who places the higher bid on item $j$ wins it. If both players bid the same amount for an item $j$, that is, $x_{1j} = x_{2j}$, a tie-breaking rule is applied. Denote $-i$ as the opponent of player $i$, where $i \in \{1, 2\}$. 
\begin{itemize}
  \item When $x_{1j} = x_{2j} = \min\{B_1, B_2, v_{1j}, v_{2j} \}$, if $\min\{B_i, v_{ij}\} > \min\{B_{-i}, v_{-ij}\}$ for some $i \in \{1, 2\}$, then player $i$ wins item $j$.
  \item In the case of all other ties, each player wins with probability $\frac{1}{2}$.
\end{itemize}
The utility of player $i$ for item $j$ is given by: 
\begin{align*}
    u_{ij}(x_{ij}, x_{-ij}) = \begin{cases}
        v_{ij} - x_{ij}, \quad & \text{if player $i$ wins item $j$}; \\
        - x_{ij}, \quad & \text{if player $i$ loses}.
    \end{cases}
\end{align*}
The total utility of player $i$ is the sum of the utilities from all items:  
\begin{align*}
    u_i(\bm{x}_i, \bm{x}_{-i}) = \sum_{j=1}^n u_{ij}(x_{ij}, x_{-ij}).
\end{align*}
A game is represented as follows:
\begin{align*}
    \mathcal{G} = \{\{1, 2\}, \{1, 2, \dots, n\}, B_1, B_2, (v_{1j})_{j=1}^n, (v_{2j})_{j=1}^n, u_1, u_2\}.
\end{align*}

A mixed strategy of player $i$ is an $n$-dimensional probability distribution over their pure strategy set $\bm{X}_i$, characterized by the cumulative distribution function $F_i(\bm{x}_i)$ and the probability density function $f_i(\bm{x}_i)$.
Given that player $-i$ follows a strategy with cumulative distribution function $F_{-i}(\bm{x}_{-i})$, when player $i$ bids a pure strategy $\bm{x}_i$, their expected utility is 
\begin{align*}
    u_i(\bm{x}_i, F_{-i}) = \mathbb{E}_{\bm{X}_{-i} \sim F_{-i}} \left[u_i(\bm{x}_i, \bm{X}_{-i})\right].
\end{align*}
Accordingly, the overall expected utility of player $i$ under the strategy $F_i$ is given by:
\begin{align*}
    \mathbb{E} [u_i(F_i, F_{-i})] = \mathbb{E}_{\bm{x}_i \sim F_i} \left[u_i(\bm{x}_i, F_{-i})\right].
\end{align*}
A strategy profile $(F_i^*, F_{-i}^*)$ is a Nash Equilibrium if and only if
\begin{align*}
    \mathbb{E} [u_i(F_i^*, F_{-i}^*)] = \max_{F_i} \mathbb{E} [u_i(F_i, F_{-i}^*)], \quad \forall i \in \{1, 2\}.
\end{align*}

Now, we provide an example to illustrate the reasoning behind the tie-breaking rule we have proposed.

\noindent\textbf{Example}:
Suppose that player 1 and player 2 have budgets of $B_1 = 0$ and $B_2 = 1$, respectively, and their valuations for the item are $v_1 = 0$ and $v_2 = 1$, respectively.
Clearly, we have $\min\{B_2, v_2\} > \min\{B_1, v_1\}$, and player 1 only bids $0$.
If ties are resolved by assigning each player a 0.5 probability of winning, player 2's utility is 0.5 when bidding 0.
However, player 2 can bid $\varepsilon \in (0, 0.5)$, increasing his utility to $1 - \varepsilon > 0.5$.
Since $\varepsilon$ can be infinitely close to 0, player 2 has no best response to player 1's strategy, leading to no Nash equilibrium under this tie-breaking rule.
By contrast, if we adopt a tie-breaking rule where lets player 2 win the item, both players bidding 0 would form a Nash equilibrium, with player 1's utility being 0 and player 2's utility being 1.

\section{Nash Equilibrium with Single Item}
In this section, we analyze the Nash equilibrium for two players in a single-item scenario. 
For simplicity, we omit the footnote $j$. 
Let $F_i(x_i)$ represent player $i$'s strategy, and let $Supp(F_i)$ denote the support of player $i$'s strategy.
Define $\overline{x}_i = \sup Supp(F_i)$ and $\underline{x}_i = \inf Supp(F_i)$ as the supremum and infimum of the support of player $i$'s strategy, respectively.
We begin by discussing the pure strategy Nash equilibrium.
Then we analyze the mixed strategy Nash equilibria, focusing on the structure of the players' strategy supports and their utilities.
Finally, we derive the Nash equilibria for two players in the single-item case.

The following Lemma provides the necessary and sufficient condition for the existence of a pure strategy Nash equilibrium.
\begin{lemma}\label{lem:pure-Nash-Equilibirum-single-item}
    Given a profile $(B_1, B_2, v_1, v_2)$, if $B_1 = B_2$ and $B_2 \leq \frac{1}{2} \min\{v_1, v_2\}$, then $(B_1, B_2)$ is the unique pure strategy Nash equilibrium; 
    otherwise, pure strategy Nash equilibrium does not exist.
\end{lemma}

When a pure strategy Nash equilibrium does not exist, we analyze the mixed strategy Nash equilibrium.
First, we characterize the supremum of the supports of the two players' strategies.
Lemma \ref{lem:support-single-battlefield-supremum} shows that the suprema of the players' strategies are identical and are determined by the minimum value among $(B_1, B_2, v_1, v_2)$.
In this paper, we denote this supremum by $L$.
\begin{lemma}\label{lem:support-single-battlefield-supremum}
    Let $L = \min\{B_1, B_2, v_1, v_2 \}$, we have $\overline{x}_1 = \overline{x}_2 = L$ in the Nash equilibrium.
\end{lemma}

According to Lemma \ref{lem:support-single-battlefield-supremum}, we know that $\overline{x}_1 = \overline{x}_2 = L = \min\{B_1, B_2, v_1, v_2 \}$.
Next, we analyze the infimum of the support of each player's strategy.
There are three possible cases: the first case occurs when the infima and suprema of both players' strategy supports are equal.
The case corresponds to Lemma \ref{lem:pure-Nash-Equilibirum-single-item}.
The second case arises when the infimum of the support of only one player's strategy is smaller than the supremum.
The case corresponds to Lemma \ref{lem:support-single-battlefield-infimum-1}.
The third case involves the infima of both players' strategy supports are smaller than their respective suprema.
The case corresponds to Lemma \ref{lem:support-single-battlefield-infimum-2}.
\begin{lemma}\label{lem:support-single-battlefield-infimum-1}
    In Nash equilibrium, if for some $i \in \{1, 2\}$, we have $\underline{x}_i < \overline{x}_i$ and $\underline{x}_{-i} = \overline{x}_{-i}$, then $Supp(F_i) = \{0, 1\}$ and $F_i(0) \leq 1 - \frac{2L}{v_{-i}}$.
    Additionally, player $i$'s utility is 0, and $-i$'s utility is at most $v_{-i} - 2L$.
\end{lemma}
\begin{lemma}\label{lem:support-single-battlefield-infimum-2}
    In Nash equilibrium, if $\underline{x}_1 < \overline{x}_1$ and $\underline{x}_{2} < \overline{x}_{2}$, then $\underline{x}_1 = \underline{x}_{2} = 0$.
\end{lemma}

When $\underline{x}_1 = \underline{x}_2 = 0$, we next show that when the bids lie within the interval $(0, L)$, both players' strategies are continuously randomized, and there are no mass points within this interval.
To establish this result, we employ a simplified version of Theorem 2 from literature \citep{baye-1996} as a lemma to support our analysis. 
This simplified version characterizes the support properties of a player's strategy in a Nash equilibrium without budget constraints.
\begin{lemma}\label{lem:support-single-battlefield-Baye-1996}
    (Simplified version of Theorem 2 from  [Baye et al., 1996])
    Two budget-unconstrained players have valuations as $v_1$ and $v_2$ for one item, $v_1 \geq v_2$, respectively.
    In Nash equilibrium, each player randomizes continuously, and there is no mass point in the interval $(0, v_2)$ for either player.
\end{lemma}

Using Lemma \ref{lem:support-single-battlefield-Baye-1996}, we derive Lemma \ref{lem:support-single-battlefield-continuously-nomass}, which states that the strategy of each player with a budget constraint is continuous on the interval $(0, L)$ and contains no mass points within this interval.
Specifically, by constructing an auction model $\hat{\mathcal{G}}$ without budget constraints, we can transform the problem with budget constraints into one without them.
By defining new strategies $\hat{F}_1$ and $\hat{F}_2$, we map the strategies from the original auction model to the budget-free model.
This mapping preserves the properties of the strategies, ensuring that the equilibrium condition still holds in the new model. 
By applying the simplified version of Lemma \ref{lem:support-single-battlefield-Baye-1996}, we conclude that the strategies of players in an auction with budget constraints retain the same properties as in the budget-free setting.

\begin{lemma}\label{lem:support-single-battlefield-continuously-nomass}
    In the Nash equilibrium, if $\underline{x}_1 = \underline{x}_{2} = 0$, then the strategy $F_i(x_i)$, $\forall i \in \{1, 2\}$, is continuously randomized over $(0, L)$, and there is no mass point within the interval $(0, L)$.
\end{lemma}

Now, we have identified all possible structures of the support of the players' strategies, namely:
\begin{itemize}
    \item $Supp(F_i) = \{L\}$, $\forall i \in \{1, 2\}$;
    \item $Supp(F_i) = \{0, L\}$ and $Supp(F_{-i}) = \{L\}$, $\exists i \in \{1, 2\}$;
    \item $\overline{Supp(F_i)} = \{x | 0 \leq x \leq L\}$, $\forall i \in \{1, 2\}$, where $\overline{Supp(F_i)}$ is the closed set of $Supp(F_i)$.
\end{itemize}

We provide an instance to show that Nash equilibrium does not exist when $B_1 = B_2$.
Specifically, when $B_1 = B_2$ and $B_1 > \frac{1}{2} \min\{v_1, v_2\}$, it is possible that the Nash equilibrium does not exist.
\begin{proposition}\label{propositon-single-item}
    When $B_1 = B_2$, there exists an instance in which the Nash equilibrium does not exist.
\end{proposition}

Let's analyze the situation where $B_1 \neq B_2$ and $\overline{Supp(F_i)} = \{x | 0 \leq x \leq L\}$, $\forall i \in \{1, 2\}$.
In this situation, we already know that each player's strategy is continuous on the interval $(0, L)$ and contains no mass points.
Now, we focus on analyzing the behavior of players at the endpoints.
Lemma \ref{lem:support-single-battlefield-endpoint-measure} characterizes the endpoint measurements and the utilities of both players.
\begin{lemma}\label{lem:support-single-battlefield-endpoint-measure}
    When $B_1 \neq B_2$, and $\underline{x}_1 = \underline{x}_{2} = 0$ in Nash equilibrium, define $s = \arg\max_{i \in \{1, 2\}} B_i$ and $w$ to be the other player.
    There are two cases in the Nash equilibrium, namely
    \begin{itemize}
        \item \textbf{Case 1:} if $v_s > L$, then the expected utility of player $s$ is  $v_s - L$, and the expected utility of player $w$ is $0$.
        The probability that player $w$ bids at 0 is $\frac{v_s - L}{v_s}$ and the probability of bidding at $L$ is 0.
        The probability that player $s$ bids at $L$ is $1 - \frac{L}{v_{w}}$ and the probability of bidding at 0 is 0.
        \item \textbf{Case 2:} if $v_s = L$, then the expected utility of player $s$ is 0, and the expected utility of player $w$ is $v_{w} - L$.
        The probability that player $s$ bids at 0 is $\frac{v_{w} - L}{v_{w}}$ and the probability of bidding at $L$ is 0.
        The probability that player $w$ bids at 0 is 0 and the probability of bidding at $L$ is 0.
    \end{itemize}
\end{lemma}

Based on Lemmas \ref{lem:pure-Nash-Equilibirum-single-item} to \ref{lem:support-single-battlefield-endpoint-measure}, we deduce Lemma \ref{lem:support-single-battlefield-nash-equilibrium-support}, which describes the possible structure of $Supp(F_i)$ in Nash equilibrium.
\begin{lemma}\label{lem:support-single-battlefield-nash-equilibrium-support}
    Given the profile $(B_1, B_2, v_1, v_2)$, if Nash equilibrium exists, we have:
    \begin{itemize}
        \item Case (1): if $B_1 = B_2$ and $B_2 < \frac{1}{2} \min\{v_1, v_2\}$, then $\underline{x}_i = \overline{x}_i$ for any $i \in \{1, 2\}$;
        \item Case (2): if $B_1 = B_{2}$ and $B_1 = \frac{1}{2} \min\{v_1, v_2\}$, let $i = \arg\min_{i' \in \{1, 2\}} v_{i'}$, then $\underline{x}_i = \overline{x}_i$, $\underline{x}_{-i} = \overline{x}_{-i}$, or $\underline{x}_i = 0$ and $\underline{x}_{-i} = \overline{x}_{-i}$;
        \item Case (3): if $B_1 \neq B_{2}$, then $\underline{x}_1 = \underline{x}_{2} = 0$.
    \end{itemize}
\end{lemma}

Based on Lemmas \ref{lem:pure-Nash-Equilibirum-single-item} to \ref{lem:support-single-battlefield-nash-equilibrium-support}, we derive the main theorem of this section, which characterizes the Nash equilibrium for two players with budget constraints in a single-item auction.
\begin{theorem}\label{thm:nash-equilibrium-single-battlefield}
    Given the profile $(B_1, B_2, v_1, v_2)$, the Nash equilibrium is as follows:
    \begin{enumerate}[label = \textbf{Case (\arabic*):}, leftmargin = 6em]
        \item When \(B_1 = B_2\) and \(B_2 < \frac{1}{2} \min\{v_1, v_2\}\), then the unique Nash equilibrium is given by the pure strategy profile \((B_1, B_2)\).
        \item When $B_1 = B_{2}$ and $B_1 = \frac{1}{2} \min\{v_1, v_2\}$, let $i = \arg\min_{i' \in \{1, 2\}} v_{i'}$, then Nash equilibrium is as follows: $Supp(F_i) = \{0, L\}$ where $F_i(0) \leq 1 - \frac{2L}{v_{-i}}$, and $Supp(F_{-i}) = \{L\}$. 
        \item When $B_1 \neq B_{2}$, define $s = \arg\max_{i \in \{1, 2\}} B_i$ and $w$ to be the another player.      
        \begin{itemize}
            \item \textbf{When \(v_s > L\):}
            \[
            F_s(x) = 
            \begin{cases}
                \frac{x}{v_w}, & x \in [0, L), \\
                1, & x = L,
            \end{cases}
            \quad
            F_w(x) = 
            \begin{cases}
                \frac{v_s - L}{v_s}, & x = 0, \\
                \frac{v_s - L + x}{v_s}, & x \in (0, L].
            \end{cases}
            \]
            \item \textbf{When \(v_s = L\):}
            \[
            F_s(x) = 
            \begin{cases}
                \frac{v_w - L}{v_w}, & x = 0, \\
                \frac{v_w - L + x}{v_w}, & x \in (0, L],
            \end{cases}
            \quad
            F_w(x) = \frac{x}{v_s}, \quad x \in [0, L].
            \]
        \end{itemize}
    \end{enumerate}
\end{theorem}

\section{Nash Equilibrium with Two Items}
In this section, we study the Nash equilibrium for two players competing over two items.
Each player's strategy is a two-dimensional joint distribution that satisfies their budget constraints.
Given the profile $(B_1, B_2, v_{11}, v_{12}, v_{21}, v_{22})$, we construct a strategy profile for both players and then verify that the constructed profile forms a Nash equilibrium.
Let $Supp(F_i)$ denote the support of player $i$'s strategy, and let $Supp(F_{ij})$ denote the support of player $i$'s strategy on item $j$.
When there are multiple items, we use the joint density function $f_i$ to represent player $i$'s strategy.

First, we focus on the case of complete symmetry.
\begin{theorem}\label{thm:two-item-nash-equilibrum-symmetric}
    Given the profile $(B_1, B_2, v_{11}, v_{12}, v_{21}, v_{22})$, if $B_1 = B_2$ and $v_{11} = v_{12} = v_{21} = v_{22}$, then
    \begin{align*}
        f_i(x_{i1}, x_{i2}) = \frac{1}{\sqrt{2} c}, \quad x_{ij} \in [0, c], x_{i2} = - x_{i1} + c, i \in \{1, 2\}, j \in \{1, 2\},
    \end{align*}
    where $c = \min\{v_{11}, B_1\}$, can form a Nash equilibrium.
\end{theorem}

According to Theorem \ref{thm:two-item-nash-equilibrum-symmetric}, we consider the following instance: $B_1 = B_2 = 1$ and $v_{11} = v_{12} = v_{21} = v_{22} = 3$.
Clearly, $\forall i \in \{1, 2\}$ and $\forall j \in \{1, 2\}$, $x_{ij} \in [0, 1]$, $x_{i2} = - x_{i1} + 1$, $f_i(x_{i1}, x_{i2}) = \frac{1}{\sqrt{2}}$ can form a Nash equilibrium.
However, the marginal distribution of player $i$ on item $j$ is $F_{ij} = x$, $x \in [0, 1]$.
By Theorem \ref{thm:nash-equilibrium-single-battlefield}, the Nash equilibrium of two players on a single item should be $(1, 1)$.
This implies that under the Nash equilibrium for two items, the marginal distribution of a player's two-dimensional joint strategy on a single item does not necessarily form a Nash equilibrium on that item.
Thus, we obtain the following corollary.
\begin{corollary}
    For each item $j$, $\forall j \in \{1, 2\}$, the strategies induced on each item $j$, $F_{ij}$, $\forall i \in \{1, 2\}$, do not necessarily constitute a Nash equilibrium.
\end{corollary}

Next, we analyze the case where $B_1 \neq B_2$.
Define $s = \arg\max_{i \in \{1, 2\}} B_i$ as the player with the larger budget, and let $w$ denote the other player.
For each item $j \in \{1, 2\}$, define $L_j = \min\{B_1, B_2, v_{1j}, v_{2j}\}$.
Without loss of generality, we assume $L_1 \geq L_2$.
According to Theorem \ref{thm:nash-equilibrium-single-battlefield}, we need to analyze four cases.

For each case, we construct a Nash equilibrium $(f_1(x_{11}, x_{12}), f_2(x_{21}, x_{22}))$.
The cases are as follows:
\begin{itemize}
    \item Case 1: $v_{s1} > L_1$ and $v_{s2} > L_2$, see Theorem \ref{thm:nash-equilibrium-two-items-case1};
    \item Case 2: $v_{s1} = L_1$ and $v_{s2} = L_2$, see Theorem \ref{thm:nash-equilibrium-two-items-case2};
    \item Case 3: $v_{s1} > L_1$ and $v_{s2} = L_2$, see Theorem \ref{thm:nash-equilibrium-two-items-case3};
    \item Case 4: $v_{s1} = L_1$ and $v_{s2} > L_2$, see Theorem \ref{thm:nash-equilibrium-two-items-case4}.
\end{itemize}
We provide a proof for Case 1, as the proofs for the other cases are similar and are thus omitted.

\begin{theorem}\label{thm:nash-equilibrium-two-items-case1}
    Given the profile $(B_1, B_2, v_{11}, v_{12}, v_{21}, v_{22})$ with $B_1 \neq B_2$, if $v_{s1} > L_1$ and $v_{s2} > L_2$, then the following strategies can form a Nash equilibrium:
    \begin{align*}
        f_s(x_{s1}, x_{s2}) = \begin{cases}
            \frac{1}{v_{w1}}, & x_{s1} \in [0, T_1], x_{s2} = L_2; \\
            \frac{1}{v_{w2}}, & x_{s1} = L_1, x_{s2} \in [0, T_2]; \\
            \frac{\frac{L_2}{v_{w2}} + \frac{L_1}{v_{w1}} - 1}{\sqrt{(L_1 - T_1)^2 + (L_2 - T_2)^2}}, & \begin{array}{l}
                x_{s1} \in (T_1, L_1), x_{s2} \in (T_2, L_2), \\
                \frac{T_2 - L_2}{L_1 - T_1} x_{s1} + T_2 - \frac{T_2 - L_2}{L_1 - T_1}L_1 = x_{s2},
            \end{array}
        \end{cases}
    \end{align*}
    where $T_1 = v_{w1} - \frac{v_{w1}}{v_{w2}} L_2$, $T_2 = v_{w2} - \frac{v_{w2}}{v_{w1}} L_1$.
    
    \begin{align*}
        f_w(x_{w1}, x_{w2}) = \begin{cases}
            \frac{1}{v_{s1}}, & x_{w1} \in [T_3, L_1], \, x_{w2} = 0; \\
            \frac{1}{v_{s2}}, & x_{w1} = 0, \, x_{w2} \in [T_4, L_2]; \vspace{3pt} \\
            \frac{\frac{L_2}{v_{s2}} + \frac{L_1}{v_{s1}} - 1}{\sqrt{T_3^2 + T_4^2}}, & x_{w1} \in (0, T_3), \, x_{w2} \in (0, T_4), \, x_{w2} = -\frac{T_4}{T_3}x_{w1} + T_4,
        \end{cases}
    \end{align*}
    where $T_3 = L_1 - \frac{v_{s2} - L_2}{v_{s2}} v_{s1}$, $T_4 = L_2 - \frac{v_{s1} - L_1}{v_{s1}} v_{s2}$.
\end{theorem}
\begin{figure}[htbp]
    \centering
    \begin{subfigure}[b]{0.35\textwidth}
        \centering
        \includegraphics[width=\textwidth]{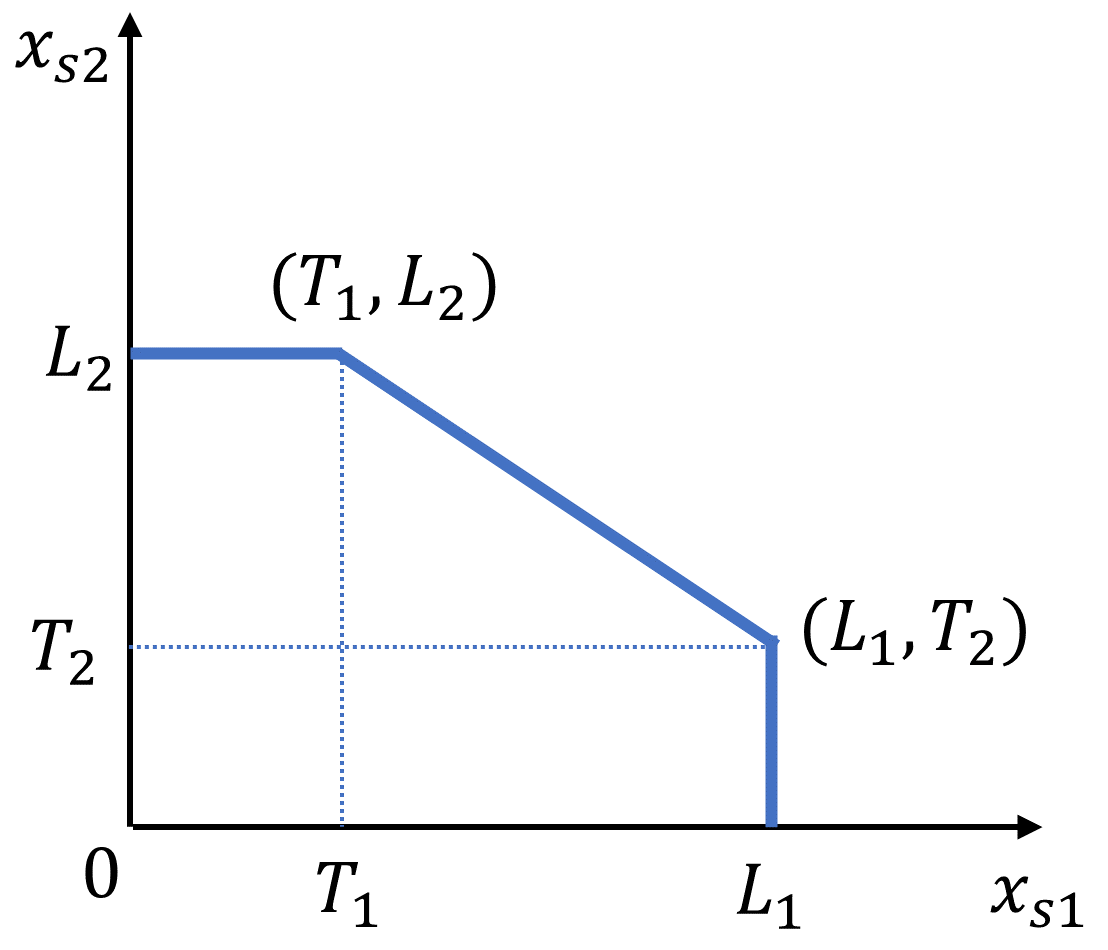}
        \caption{$Supp(F_s)$}
    \end{subfigure}
    \begin{subfigure}[b]{0.35\textwidth}
        \centering
        \includegraphics[width=\textwidth]{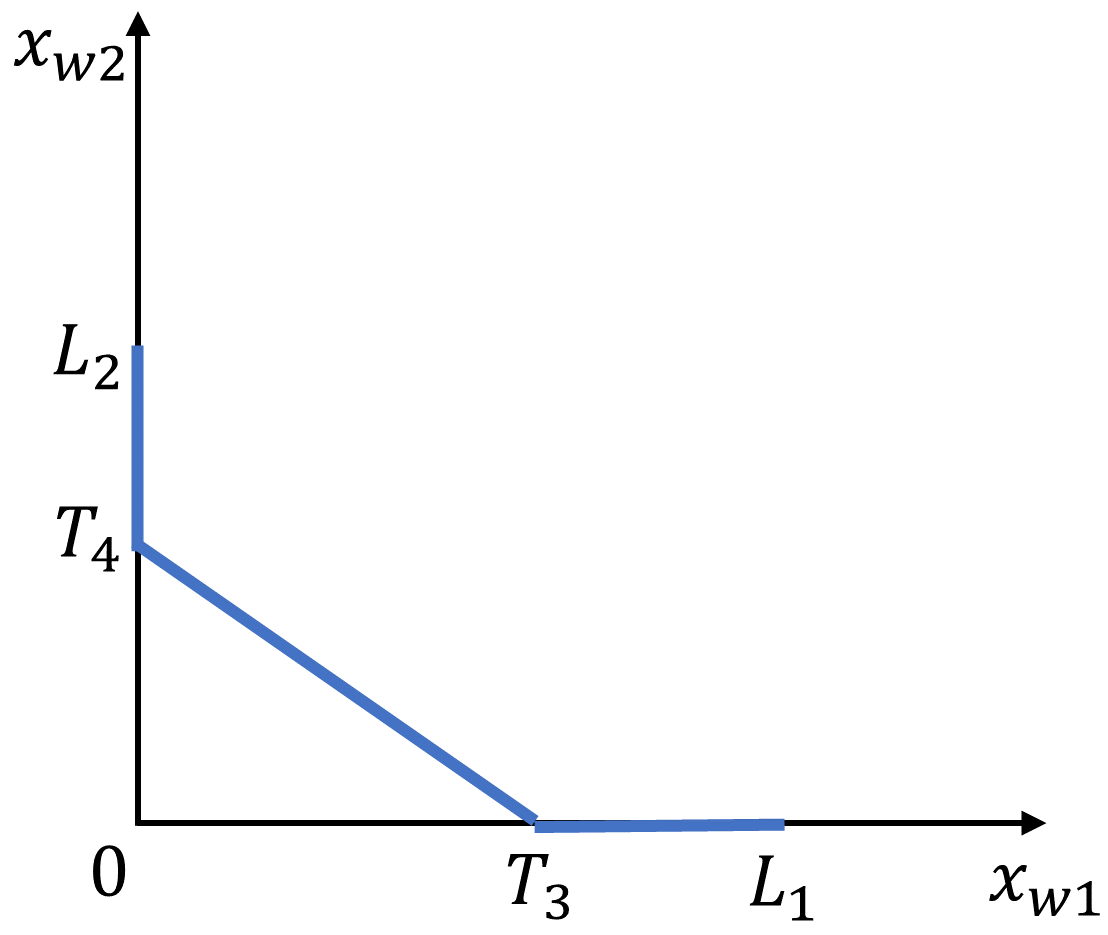}
        \caption{$Supp(F_w)$}
    \end{subfigure}
    \caption{The blue line represents the support of players' strategies for Case 1.}
    \label{fig:case1}
\end{figure}
Figure \ref{fig:case1} illustrates the structure of the support of both players' strategies in Case 1.
We observe that both players' strategies still form a Nash equilibrium within a single item.
Player $s$ forms a mass point at both $x_{s1} = L_1$ and $x_{s2} = L_2$, respectively.
Meanwhile, player $w$ forms a mass point at both $x_{w1} = 0$ and $x_{w2} = 0$, respectively.
\begin{theorem}\label{thm:nash-equilibrium-two-items-case2}
    Given the profile $(B_1, B_2, v_{11}, v_{12}, v_{21}, v_{22})$ with $B_1 \neq B_2$, if $v_{s1} = L_1$ and $v_{s2} = L_2$, then the following strategies can form a Nash equilibrium:
    \begin{align*}
        f_s(x_{s1}, x_{s2}) = \begin{cases}
            \frac{\frac{L_1}{v_{w1}} + \frac{L_2}{v_{w2}} - 1}{\sqrt{T_2^2 + T_1^2}}, &x_{s1} \in [0, T_1), \, x_{s2} \in [0, T_2), \, x_{s2} = -\frac{T_2}{T_1} x_{s1} + T_2; \\
            \frac{1}{v_{w1}}, &x_{s1} \in [T_1, L_1], \, x_{s2} = 0; \\
            \frac{1}{v_{w2}}, &x_{s1} = 0, \, x_{s2} \in [T_2, L_2],
        \end{cases}
    \end{align*}
    where $T_1 = L_1 - v_{w1} + \frac{L_2}{v_{w2}} v_{w1}$, $T_2 = L_2 - v_{w2} + \frac{L_1}{v_{w1}} v_{w2}$.
    \begin{align*}
        f_w(x_{w1}, x_{w2}) = \frac{1}{\sqrt{L_1^2 + L_2^2}}, \quad x_{w1} \in [0, L_1], \, x_{w2} \in [0, L_2], \, x_{w2} = -\frac{L_2}{L_1} x_{w1} + L_2.
    \end{align*}
\end{theorem}

\begin{figure}[htbp]
    \centering
    \begin{subfigure}[b]{0.35\textwidth}
        \centering
        \includegraphics[width=\textwidth]{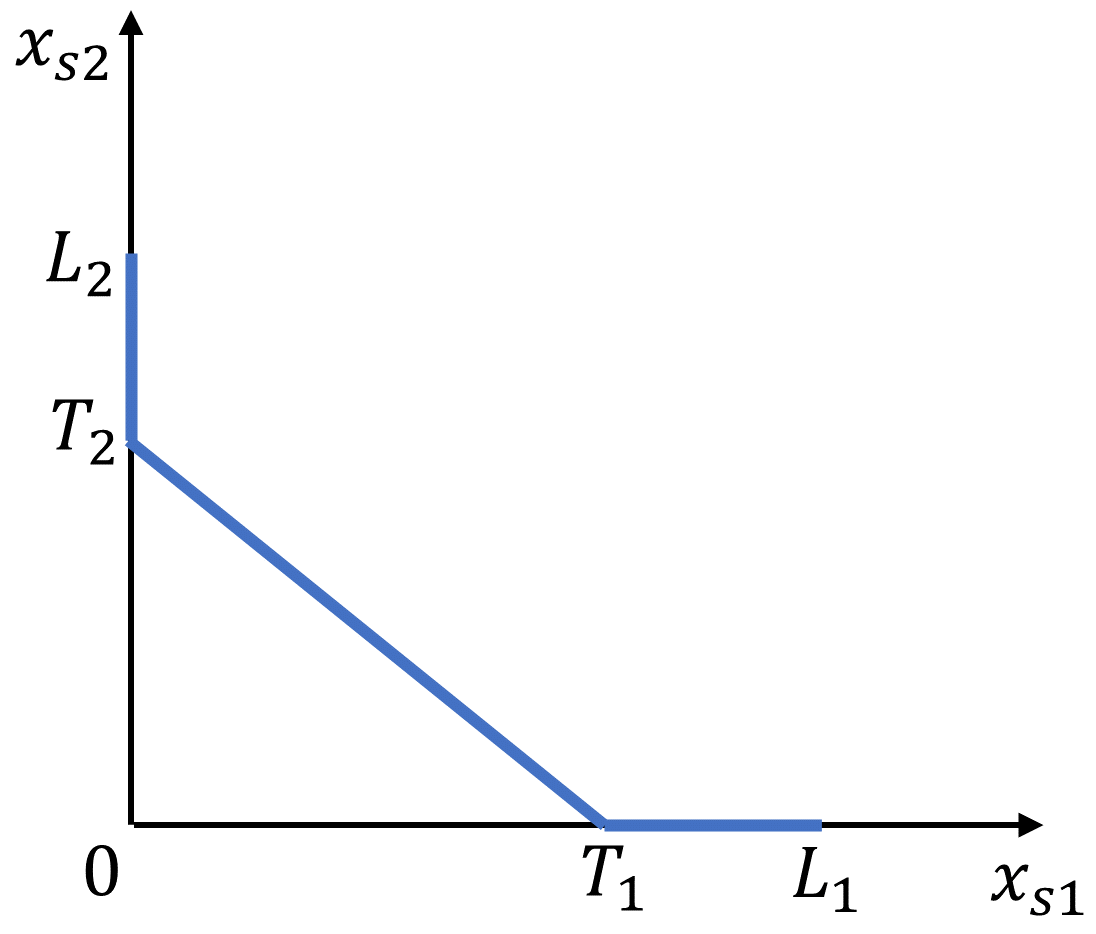}
        \caption{$Supp(F_s)$}
    \end{subfigure}
    \begin{subfigure}[b]{0.35\textwidth}
        \centering
        \includegraphics[width=\textwidth]{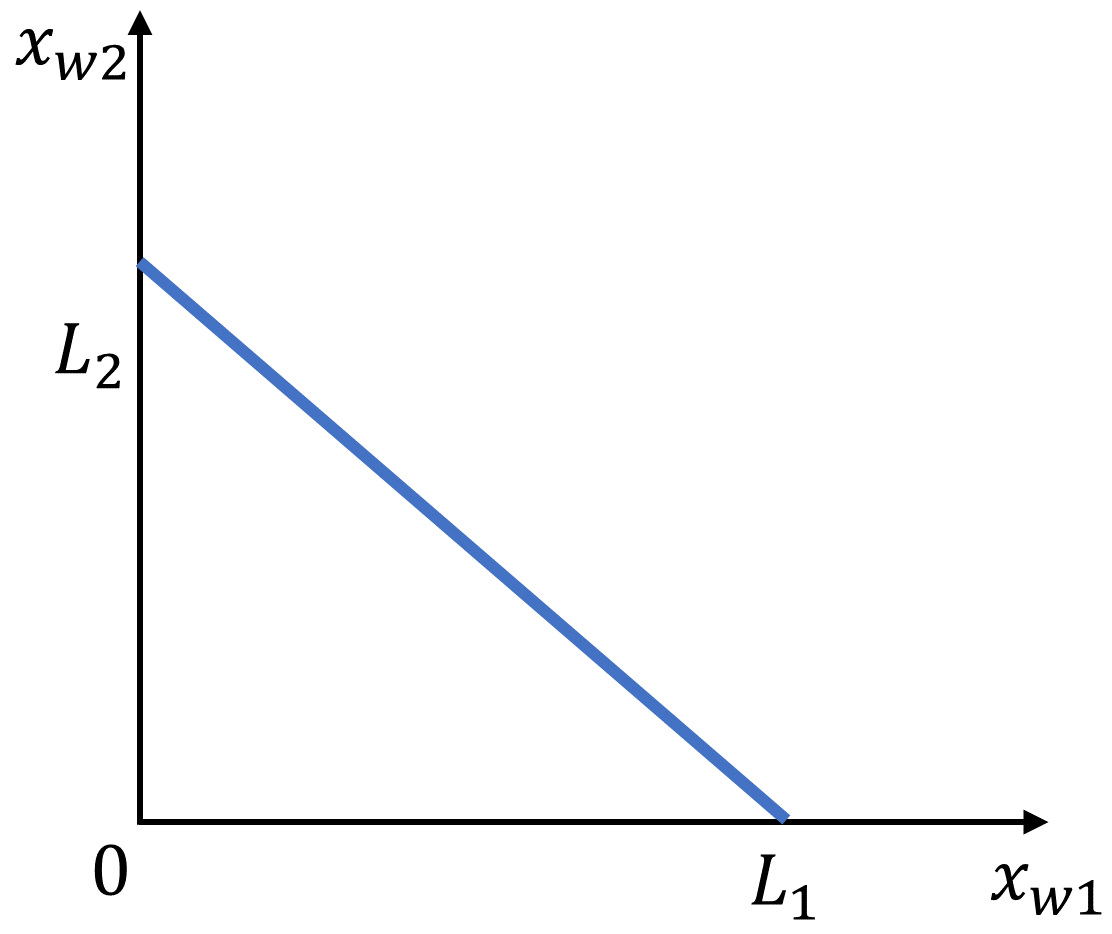}
        \caption{$Supp(F_w)$}
    \end{subfigure}
    \caption{The blue line represents the support of players' strategies for Case 2.}
    \label{fig:case2}
\end{figure}
Figure \ref{fig:case2} illustrates the structure of the support of both players' strategies in Case 2.
We observe that both players' strategies also form a Nash equilibrium within a single item.
Since player $s$ has a low valuation for every item, he forms a mass point at both $x_{s1} = 0$ and $x_{s2} = 0$, respectively.
In contrast, player $w$ has no mass points in his support.

\begin{theorem}\label{thm:nash-equilibrium-two-items-case3}
    Given the profile $(B_1, B_2, v_{11}, v_{12}, v_{21}, v_{22})$ with $B_1 \neq B_2$, if $v_{s1} > L_1$ and $v_{s2} = L_2$, then the following strategies form a Nash equilibrium:
    \begin{align*}
        f_s(x_{s1}, x_{s2}) = \begin{cases}
            \frac{1}{v_{w2}}, &x_{s1} = L_1, \, x_{s2} \in (0, T_2]; \\
            \frac{\frac{L_1}{v_{w1}}}{\sqrt{L_1^2 + (L_2 - T_2)^2}}, & \begin{array}{l}
                x_{s1} \in [0, L_1), \, x_{s2} \in (T_2, L_2], \\ 
                x_{s2} = -\frac{L_2 - T_2}{L_1} x_{s1} + L_2,
            \end{array}
        \end{cases}
    \end{align*}
    where $T_2 = -\frac{L_1}{v_{w1}} v_{w2} + L_2$.
    When $x_{s1} = L_1, \, x_{s2} = 0$, there is a mass point, the probability value is $1 - \frac{L_2}{v_{w2}}$.
    \begin{align*}
        f_w(x_{w1}, x_{w2}) = \begin{cases}
            \frac{1}{v_{s2}}, &x_{w1} = 0, \, x_{w2} \in [T_4, L_2]; \\
            \frac{\frac{L_1}{v_{s1}}}{\sqrt{L_1^2 + T_4^2}}, &x_{w1} \in (0, L_1], \, x_{w2} \in [0, T_4), \, x_{w2} = -\frac{T_4}{L_1} x_{w1} + T_4,
        \end{cases}
    \end{align*}
    where $T_4 = L_2 - \frac{v_{s1} - L_1}{v_{s1}} v_{s2}$.
\end{theorem}

Theorem \ref{thm:nash-equilibrium-two-items-case4}, which corresponds to Case 4, will be moved to Appendix due to space limitations.

\section{Nash Equilibrium with Three Items}
In this section, we study the Nash equilibrium for three items.
When analyzing the two-item case, the values of the items are asymmetric between players.
However, when there are three items, analyzing the Nash equilibrium becomes highly complex.
To simplify the analysis, we consider the symmetric valuation case, where the value of each item is the same for both players, namely $v_{1j} = v_{2j}$ for $j \in \{1, 2, 3\}$.
We construct a strategy for each player and, in the end, verify that the constructed strategies form a Nash equilibrium.

For three items, let $v_j$ denote the valuation of item $j$, without loss of generality, we assume $v_1 \geq v_2 \geq v_3$ and $B_1 \geq B_2$.
The following theorem provides a Nash equilibrium when $B_i \geq \max\{\frac{v_1 + v_2 + v_3}{2}, v_1\}$, $\forall i \in \{1, 2\}$.
\begin{theorem}\label{thm:nash-equilibrium-three-items}
    Given the profile $(B_1, B_2, v_1, v_2, v_3)$, if $B_i \geq \max\{\frac{v_1 + v_2 + v_3}{2}, v_1\}$, $\forall i \in \{1, 2\}$, then we have two cases:
    \begin{itemize}
        \item Case 1: $\frac{v_1 + v_2 + v_3}{2} > v_1$.
        Let $z = \frac{v_1 + v_2 + v_3}{2}$, $A = (v_1, 0, z - v_1)$, $B = (z - v_2, v_2, 0)$, $C = (0, z - v_3, v_3)$.
        Let $L_{AB}$, $L_{BC}$, and $L_{CA}$ denote the line segments with endpoints $A$ and $B$, $B$ and $C$, and $C$ and $A$, respectively.
        Let $|AB|$, $|BC|$, and $|CA|$ denote the lengths of the line segments $L_{AB}$, $L_{BC}$, and $L_{CA}$, respectively.
        Then for player $i \in \{1, 2\}$,
        \begin{align*}
            f_i(x_{i1}, x_{i2}, x_{i3}) = \begin{cases}
                \frac{P_{AB}}{|AB|}, \quad  & if \, (x_{i1}, x_{i2}, x_{i3}) \text{ is in } L_{AB}; \\
                \frac{P_{BC}}{|BC|}, & if \, (x_{i1}, x_{i2}, x_{i3})  \text{ is in } L_{BC}; \\
                \frac{P_{CA}}{|CA|}, & if \, (x_{i1}, x_{i2}, x_{i3}) \text{ is in } L_{CA},
            \end{cases}
        \end{align*}
        where $P_{AB} = \frac{1}{\frac{z - v_3}{z - v_2} + 1 + \frac{z - v_3}{z - v_1}} \cdot \frac{z - v_3}{z - v_2}$, $P_{BC} = \frac{1}{\frac{z - v_3}{z - v_2} + 1 + \frac{z - v_3}{z - v_1}}$, $P_{CA} = \frac{1}{\frac{z - v_3}{z - v_2} + 1 + \frac{z - v_3}{z - v_1}} \cdot \frac{z - v_3}{z - v_1}$, can form a Nash equilibrium.
        \item Case 2: $\frac{v_1 + v_2 + v_3}{2} \leq v_1$.
        Define \( A = (v_1, 0, 0) \) and \( B = (0, v_2, v_3) \). Let \( f_i(x_{i1}, x_{i2}, x_{i3}) \) be the uniform joint density function on the line segment \( AB \), i.e., \( f_i(x_{i1}, x_{i2}, x_{i3}) = \frac{1}{|AB|} \). 
        Then, \( f_i(x_{i1}, x_{i2}, x_{i3}) \) for player \( i \in \{1, 2\} \) can form a Nash equilibrium.
    \end{itemize}
\end{theorem}

\begin{figure}[h]
\centering
\begin{tikzpicture}[scale=0.7, transform shape]
    \begin{axis}
    [
        name=left, 
        at={(0,0)}, 
        view={130}{30}, 
        axis lines=center,
        xlabel={$x_{i1}$}, ylabel={$x_{i2}$}, zlabel={$x_{i3}$},
        xtick=\empty, ytick=\empty, ztick=\empty,
        grid=major,
        zmax = 9,
        xmax = 12,
        ymax = 10
    ]
        \addplot3[only marks, mark=*, mark size=0pt, blue] coordinates {(10,0,2)};
        \addplot3[only marks, mark=*, mark size=0pt, blue] coordinates {(4,8,0)};
        \addplot3[only marks, mark=*, mark size=0pt, blue] coordinates {(0,6,6)};
    
        \addplot3[line width=2pt, blue] coordinates {(10,0,2) (4,8,0)};
        \addplot3[line width=2pt, blue] coordinates {(4,8,0) (0,6,6)};
        \addplot3[line width=2pt, blue] coordinates {(0,6,6) (10,0,2)};
    
        \node[anchor=west] at (axis cs:11.5, 0, 3.5) {A};
        \node[anchor=east] at (axis cs:5.2, 9.5, 0) {B};
        \node[anchor=north] at (axis cs:0, 5.8, 7.5) {C};
        \node[anchor=north] at (axis cs:0, 0, 0) {0};
    \end{axis}
    \node[anchor=north] at (3.5,0) {\large (a) Support for Case 1};

    \begin{axis}
    [
        name = right, 
        at = {(7.5cm,0)}, 
        view={130}{30}, 
        axis lines=center,
        xlabel={$x_{i1}$}, ylabel={$x_{i2}$}, zlabel={$x_{i3}$},
        xtick=\empty, ytick=\empty, ztick=\empty,
        grid=major,
        zmax = 9,
        xmax = 12,
        ymax = 10
    ]
        \addplot3[only marks, mark=*, mark size=0pt, red] coordinates {(10,0,0)};
        \addplot3[only marks, mark=*, mark size=0pt, red] coordinates {(0,4,4)};

        \addplot3[line width=2pt, blue] coordinates {(10,0,0) (0,4,4)};
    
        \node[anchor=south west] at (axis cs:13.5, 2, 0.5) {A}; 
        \node[anchor=north east] at (axis cs:0, 5, 4.5) {B};
        \node[anchor=north] at (axis cs:0, 0, 0) {0};
    \end{axis}
    \node[anchor=north] at (11,0) {\large (b) Support for Case 2};
\end{tikzpicture}
\caption{The support of players' strategies for three items}
\label{fig:three-item}
\end{figure}
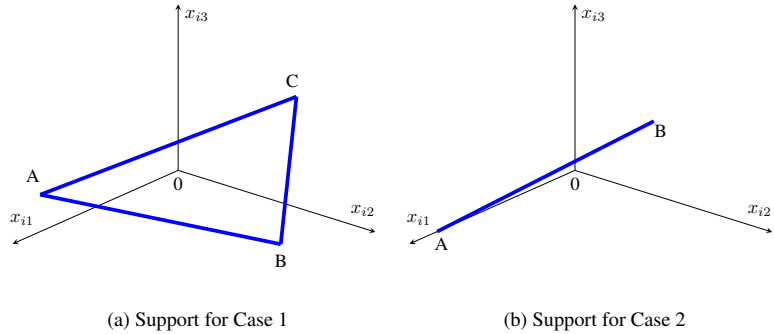

Figure \ref{fig:three-item} illustrates the support of both players' strategies.
It can be verified that the three-dimensional joint density function we constructed has uniform marginal density functions in each dimension.
From the perspective of a single item, both players still form a Nash equilibrium on each individual item.

\section{Conclusion and Future Works}
We investigate the Nash equilibrium in an all-pay auction where two players with budget constraints compete for multiple items. 
We find that when players have budget constraints, a Nash equilibrium does not always exist, and for a single item, the Nash equilibrium may not be unique.
For multiple items, we construct Nash equilibrium strategies for the two players.
However, we have only discussed cases with up to three items.
For a larger number of items, the existence of a Nash equilibrium remains an open question. 
Additionally, for the three-item case, we have provided Nash equilibrium only for certain special cases.
In the future, analyzing Nash equilibrium in more general settings for the three-item case will be our research focus.

\clearpage
\bibliographystyle{named}
\bibliography{reference}

\begin{thebibliography}{}

\bibitem[\protect\citeauthoryear{Avni \bgroup \em et al.\egroup }{2020}]{avni-2020}
Guy Avni, Rasmus Ibsen-Jensen, and Josef Tkadlec.
\newblock All-pay bidding games on graphs.
\newblock In {\em Proceedings of the AAAI Conference on Artificial Intelligence}, volume~34, pages 1798--1805, 2020.

\bibitem[\protect\citeauthoryear{Avni \bgroup \em et al.\egroup }{2021}]{avni-2021}
Guy Avni, Ism{\"a}el Jecker, and {\DJ}or{\dj}e {\v{Z}}ikeli{\'c}.
\newblock Infinite-duration all-pay bidding games.
\newblock In {\em Proceedings of the 2021 ACM-SIAM Symposium on Discrete Algorithms (SODA)}, pages 617--636. SIAM, 2021.

\bibitem[\protect\citeauthoryear{Barut and Kovenock}{1998}]{barut-1998}
Yasar Barut and Dan Kovenock.
\newblock The symmetric multiple prize all-pay auction with complete information.
\newblock {\em European Journal of Political Economy}, 14(4):627--644, 1998.

\bibitem[\protect\citeauthoryear{Baye \bgroup \em et al.\egroup }{1996}]{baye-1996}
Michael~R Baye, Dan Kovenock, and Casper~G De~Vries.
\newblock The all-pay auction with complete information.
\newblock {\em Economic Theory}, 8(2):291--305, 1996.

\bibitem[\protect\citeauthoryear{Boix-Adser{\`a} \bgroup \em et al.\egroup }{2020}]{boix-2020}
Enric Boix-Adser{\`a}, Benjamin~L Edelman, and Siddhartha Jayanti.
\newblock The multiplayer colonel blotto game.
\newblock In {\em Proceedings of the 21st ACM Conference on Economics and Computation}, pages 47--48, 2020.

\bibitem[\protect\citeauthoryear{Che and Gale}{1996}]{che-1996}
Yeon-Koo Che and Ian Gale.
\newblock Expected revenue of all-pay auctions and first-price sealed-bid auctions with budget constraints.
\newblock {\em Economics Letters}, 50(3):373--379, 1996.

\bibitem[\protect\citeauthoryear{Dechenaux \bgroup \em et al.\egroup }{2015}]{dechenaux-2015}
Emmanuel Dechenaux, Dan Kovenock, and Roman~M Sheremeta.
\newblock A survey of experimental research on contests, all-pay auctions and tournaments.
\newblock {\em Experimental Economics}, 18:609--669, 2015.

\bibitem[\protect\citeauthoryear{Dekel \bgroup \em et al.\egroup }{2007}]{dekel-2007}
Eddie Dekel, Matthew~O Jackson, and Asher Wolinsky.
\newblock Jump bidding and budget constraints in all-pay auctions and wars of attrition.
\newblock Technical report, Discussion Paper, 2007.

\bibitem[\protect\citeauthoryear{Dulleck \bgroup \em et al.\egroup }{2006}]{dulleck-2006}
Uwe Dulleck, Paul Frijters, and Konrad Podczeck.
\newblock All-pay auctions with budget constraints and fair insurance.
\newblock Technical report, Working Paper, 2006.

\bibitem[\protect\citeauthoryear{Dziubi{\'n}ski and Jahn}{2023}]{dziubinski-2023}
Marcin Dziubi{\'n}ski and Krzysztof Jahn.
\newblock Discrete two player all-pay auction with complete information.
\newblock In {\em Proceedings of the Thirty-Second International Joint Conference on Artificial Intelligence}, pages 2659--2666, 2023.

\bibitem[\protect\citeauthoryear{Hillman and Riley}{1989}]{hillman-1989}
Arye~L Hillman and John~G Riley.
\newblock Politically contestable rents and transfers.
\newblock {\em Economics \& Politics}, 1(1):17--39, 1989.

\bibitem[\protect\citeauthoryear{Hwang \bgroup \em et al.\egroup }{2023}]{hwang-2023}
Sung-Ha Hwang, Youngwoo Koh, and Jingfeng Lu.
\newblock Constrained contests with a continuum of battles.
\newblock {\em Games and Economic Behavior}, 142:992--1011, 2023.

\bibitem[\protect\citeauthoryear{Kovenock and Roberson}{2021}]{kovenock-2021}
Dan Kovenock and Brian Roberson.
\newblock Generalizations of the general lotto and colonel blotto games.
\newblock {\em Economic Theory}, 71:997--1032, 2021.

\bibitem[\protect\citeauthoryear{Nalebuff and Stiglitz}{1983}]{nalebuff-1983}
Barry~J Nalebuff and Joseph~E Stiglitz.
\newblock Prizes and incentives: towards a general theory of compensation and competition.
\newblock {\em The Bell Journal of Economics}, pages 21--43, 1983.

\bibitem[\protect\citeauthoryear{Roberson and Kvasov}{2012}]{roberson-2012}
Brian Roberson and Dmitriy Kvasov.
\newblock The non-constant-sum colonel blotto game.
\newblock {\em Economic Theory}, 51:397--433, 2012.

\bibitem[\protect\citeauthoryear{Roberson}{2006}]{roberson-2006}
Brian Roberson.
\newblock The colonel blotto game.
\newblock {\em Economic Theory}, 29(1):1--24, 2006.

\end{thebibliography}
\clearpage

\noindent {\textbf{\Large APPENDIX}}

\vspace{10pt}
\noindent {\textbf{\large A \quad Missing proofs in Section 3}}

\vspace{5pt}
\noindent {\textbf{Proof of Lemma \ref{lem:pure-Nash-Equilibirum-single-item}}}
\begin{proof}
    First, we prove that $(B_1, B_2)$ is a pure strategy Nash equilibrium when this condition is met.
    When player $-i$ bids $x_{-i} = B_{-i}$, we have
    \begin{align*}
        u_i(B_i, B_{-i}) = \frac{1}{2} v_i - B_i \geq 0 \geq u_i(x_i, B_{-i}), \quad \forall x_i \in [0, B_i),
    \end{align*}
    which means that player $i$'s utility will not increase if his bid deviates from $B_i$ to $x_i \in [0, B_i)$.
    Therefore, we can easily verify that $(B_1, B_2)$ is a pure strategy Nash equilibrium.

    Next, we prove that there is no other pure strategy Nash equilibrium besides $(B_1, B_2)$.
    When player $-i$ bids $x_{-i} \in [0, B_{-i})$, for any bid $x_i$, player $i$ has motivation to deviate from $x_i$. 
    To discuss in detail, if player $i$ bids $x_{i} \in [0, x_{-i}]$, we have $u_{i}(x_{i}, x_{-i}) < u_{i}(x_{-i} + \varepsilon, x_{-i})$ as long as $0 < \varepsilon < B_{i} - x_{-i}$; 
    if player $i$ bids $x_{i} \in (x_{-i}, B_i]$, we have $u_{i}(x_{i}, x_{-i}) < u_{i}(x_{i} - \varepsilon, x_{-i})$ as long as $0 < \varepsilon < x_{i} - x_{-i}$. 
    Thus, there exists no pure strategy Nash equilibrium when player $i$ bids $x_{i} \in [0, B_{i})$.
    Therefore, $(B_1, B_2)$ is unique pure strategy Nash equilibrium.
    
    Finally, we prove that there is no pure strategy Nash equilibrium when this condition is not met.
    If $B_1 \neq B_2$, or $B_1 = B_2$ and $B_2 > \frac{1}{2} \min\{v_1, v_2\}$, let $R_1 = \min\{v_1, B_1\}$, $R_2 = \min\{v_2, B_2\}$, and $i^* = \arg\max_{i \in \{1, 2\}} R_i$, so $R_{i^*} \geq R_{-i^*}$.
    We here discuss two cases.
    \begin{enumerate}[label = Case (\arabic*), left = 1.5em]
        \item $R_{i^*} > R_{-i^*}$. 
        The bid of player $-i^*$ is $x_{-i^*} \in [0, R_{-i^*}]$, we discuss the following situations.
        \begin{itemize}
            \item Player $-i^*$ bids 0. 
            We have that $u_{i^*}(\varepsilon, 0) = v_{i^*} - \varepsilon > u_{i^*}(0, 0) = \frac{1}{2} v_{i^*}$ for sufficiently small $\varepsilon > 0$. 
            However, the $\varepsilon > 0$ can be infinitely small, the best response for player $i^*$ does not exist. 
            Therefore, a bid of 0 by player $-i^*$ cannot constitute a pure Nash equilibrium.
            \item Player $-i^*$ bids $x_{-i^*} \in (0, R_{-i^*})$.
            For $\forall x_{i^*} \in [0, x_{-i^*})$, we have $u_{i^*}(x_{i^*}, x_{-i^*}) = 0 - x_{i^*} \leq 0$, which means that if the bid of player $i^*$ is less than that of player $-i^*$, player $i^*$ will not get positive utility.
            For $x_{i^*} = x_{-i^*}$, we have $u_{i^*}(x_{i^*}, x_{-i^*}) = \frac{1}{2} v_{i^*} - x_{i^*} < v_{i^*} - (x_{i^*} + \varepsilon) = u_{i^*}(x_{i^*} + \varepsilon, x_{-i^*})$ for sufficiently small $\varepsilon > 0$.
            Note that $x_{-i^*} < R_{-i^*} < R_{i^*} \leq v_{i^*}$, therefore we have $u_{i^*}(x_{i^*} + \varepsilon, x_{-i^*}) > 0$.
            We can derive that $u_{i^*}(x_{i^*}, x_{-i^*}) < u_{i^*}(x_{-i^*} + \varepsilon, x_{-i^*})$ for $\forall x_{i^*} \in [0, x_{-i^*})$ and $u_{i^*}(x_{-i^*}, x_{-i^*}) < u_{i^*}(x_{-i^*} + \varepsilon, x_{-i^*})$.
            However, the $\varepsilon > 0$ can be infinitely small, the best response for player $i^*$ does not exist.
            Therefore, a bid of $x_{-i^*} \in (0, R_{-i^*}]$ by player $-i^*$ cannot constitute a pure Nash equilibrium.
            \item Player $-i^*$ bids $x_{-i^*} = R_{-i^*}$.
            Player $i^*$'s best response is to bid $R_{-i^*}$. 
            To discuss in detail, $u_{i^*}(R_{-i^*}, R_{-i^*}) > 0 \geq u_{i^*}(x_{i^*}, R_{-i^*})$ for any $x_{i^*} \in [0, R_{-i^*})$, $u_{i^*}(R_{-i^*}, R_{-i^*}) = v_{i^*} - R_{-i^*} > v_{i^*} - x_{i^*} =  u_{i^*}(x_{i^*}, R_{-i^*})$ for any $x_{i^*} \in (R_{-i^*}, R_{i^*}]$. 
            However, if Player $i^*$ bid $R_{-i^*}$, $u_{-i^*}(R_{-i^*}, 0) = 0 > u_{-i^*}(R_{-i^*}, R_{-i^*})$. Therefore, there is no pure Nash equilibrium in this case.
        \end{itemize}
        \item $R_{i^*} = R_{-i^*}$.
        We discuss the following situations.
        \begin{itemize}
            \item $R_{i^*} = v_{i^*}$ or $R_{-i^*} = v_{-i^*}$.
            We only discuss the situation where $R_{i^*} = v_{i^*}$; the situation where $R_{-i^*} = v_{-i^*}$ is similar and thus omitted.

            When $x_{-i^*} \in [0, R_{-i^*})$, if player $i^*$ bids $x_{i^*} \in [0, x_{-i^*})$, we have $u_{i^*}(x_{i^*}, x_{-i^*}) < u_{i^*}(x_{-i^*} + \varepsilon, x_{-i^*})$ as long as $\varepsilon < R_{-i^*} - x_{-i^*}$;
            if player $i^*$ bids $x_{i^*} = x_{-i^*}$, we have $u_{i^*}(x_{i^*}, x_{-i^*}) < u_{i^*}(x_{i^*} + \varepsilon, x_{-i^*})$ as long as $\varepsilon < R_{-i^*} - x_{i^*}$;
            if player $i^*$ bids $x_{i^*} > x_{-i^*}$, we have $u_{i^*}(x_{i^*}, x_{-i^*}) < u_{i^*}(x_{i^*} - \varepsilon, x_{-i^*})$ as long as $\varepsilon < \frac{1}{2} (x_{i^*} - x_{-i^*})$.
            Therefore, when player $-i^*$ bids $x_{-i^*} \in [0, R_{-i^*})$, there is no pure strategy Nash equilibrium.
            
            When $x_{-i^*} = R_{-i^*}$, we can deduce that if player $i^*$ bids 0, $u_{i^*}(0, R_{-i^*}) = 0$; 
            if player $i^*$ bids $x_{i^*} \in (0, R_{-i^*})$, $u_{i^*}(x_{i^*}, R_{-i^*}) < 0$; 
            if player $i^*$ bids $x_{i^*} = R_{-i^*}$, there are 3 cases: if $B_{i^*} = B_{-i^*}$, then $u_{i^*}(R_{-i^*}, R_{-i^*}) = \frac{1}{2} v_{i^*} - R_{-i^*} < 0$; if  $B_{i^*} > B_{-i^*}$, then $u_{i^*}(R_{-i^*}, R_{-i^*}) = 0$; if $B_{i^*} < B_{-i^*}$, then $u_{i^*}(R_{-i^*}, R_{-i^*}) < 0$.
            Therefore, if $B_{i^*} \leq B_{-i^*}$, then player $i^*$ can only bid 0; if $B_{i^*} > B_{-i^*}$, then player $i^*$ can bid 0 or $v_{i^*}$.
            
            However, if player $i^*$ bids $x_{i^*} = 0$, we have $u_{-i^*}(0, 0) = \frac{1}{2} v_{-i^*} < v_{-i^*} - \varepsilon = u_{-i^*}(0, \varepsilon)$ for sufficiently small $\varepsilon > 0$.
            Since $\varepsilon$ can be arbitrarily close to 0, player $-i^*$ does not have a best response to player $i^*$, so a pure strategy Nash equilibrium does not exist.
            If $B_{i^*} > B_{-i^*}$ and player $i^*$ bids $x_{i^*} = v_{i^*}$, then player $-i^*$'s best response is to bid 0, which is different from $R_{-i^*}$. Therefore, there is no pure Nash equilibrium.
            \item $R_{i^*} = B_{i^*}$ and $R_{-i^*} = B_{-i^*}$.
            In this situation, we can know that $B_{i^*} = B_{-i^*} \leq \min\{v_1, v_2\}$.
            We have proven that when $B_{i^*} = B_{-i^*} \leq \frac{1}{2} \min\{v_1, v_2\}$, $(B_{i^*}, B_{-i^*})$ is a pure strategy Nash equilibrium.
            Now we argue the situation where $\frac{1}{2} \min\{v_1, v_2\} < B_{i^*} = B_{-i^*} \leq \min\{v_1, v_2\}$.
            Similar to the above explanation, when $x_{-i^*} \in [0, R_{-i^*})$, there is no pure strategy Nash equilibrium.
            When $x_{-i^*} = R_{-i^*}$, if player $i^*$ bids 0, $u_{i^*}(0, R_{-i^*}) = 0$; 
            if player $i^*$ bids $x_{i^*} \in (0, R_{-i^*})$, $u_{i^*}(x_{i^*}, R_{-i^*}) < 0$; 
            if player $i^*$ bids $x_{i^*} = R_{-i^*}$, let $\hat{i} = \arg \min_{i \in \{1,2\}} \{v_i\}$, $u_{\hat{i}}(R_{\hat{i}}, R_{-\hat{i}}) = \frac{1}{2} v_{\hat{i}} - R_{\hat{i}} < 0$.
            This implies that there will always be a player who gets a higher utility by bidding 0 compared to bidding other values, while the other player will bid a sufficiently small $\varepsilon > 0$. 
            Clearly, this cannot constitute a pure strategy Nash equilibrium.
        \end{itemize}
    \end{enumerate}
\end{proof}

\vspace{5pt}
\noindent {\textbf{Proof of Lemma \ref{lem:support-single-battlefield-supremum}}}
\begin{proof}
    We first demonstrate that $\overline{x}_1 = \overline{x}_2$.
    We prove it by contradiction. 
    Without loss of generality, assume $\overline{x}_1 > \overline{x}_2$.
    For player 1, within the interval $(\overline{x}_2, \overline{x}_1]$, there must exist a bid $x \in Supp(F_1)$. 
    At this bid, we can derive that the utility of player 1 is $u_1(x, F_2) = v_1 - x$.
    Consider another bid $x' \in (\overline{x}_2, \overline{x}_1)$ and $x' < x$, if player 1 bids $x'$, we have that $u_1(x', F_2) = v_1 - x'$.
    Therefore, $u_1(x', F_2) > u_1(x, F_2)$, which means that if player 1's bid deviates from $x$ to $x'$, his utility increases, implying that $x$ is not in his strategy's support, which is a contradiction.
    
    We next prove that $L = \overline{x}_1 = \overline{x}_2$.
    Without loss of generality, we assume $v_1 \geq v_2$.
    We need to discuss four cases.
    \begin{enumerate}[label = Case (\arabic*), left = 1.5em]
        \item $B_1 \geq v_2$ and $B_2 \geq v_2$.
        We have $L = \min\{B_1, B_2, v_1, v_2 \} = v_2$ because of $v_1 \geq v_2$.
        If player 2's bid exceeds $v_2$, then his utility will definitely be negative. 
        Therefore, player 2's bid will not exceed $v_2$.
        Player 1's bid will not exceed $v_2$ either.
        If $\overline{x}_1 = \overline{x}_2 < L$, for player 1, $\lim_{x_1 \rightarrow \overline{x}_1^-} u_1(x_1, F_2) \leq v_1 - \overline{x}_1 = \lim_{x_1 \rightarrow \overline{x}_1^+} u_1(x_1, F_2)$, which contradicts the fact that $\overline{x}_1$ is the supremum of $Supp(F_1)$.
        Therefore, we conclude that $\overline{x}_1 = \overline{x}_2 = L$.
        \item $B_2 < v_2$ or $B_1 < v_2$. We prove it by contradiction.
        Let $L = B_i$, since player $i$ can't bid more than $L$, suppose the common supremum is $T < L$.
        For any $i$, consider player $i$'s utility when bidding out of $Supp(F_i)$, we have
        \begin{align*}
            \lim_{\varepsilon \rightarrow 0^+} u_i(T + \varepsilon, F_{-i}) = \lim_{x_i \rightarrow T^+} u_i(x_i, F_{-i}) = v_i - T > 0.
        \end{align*}
        This implies that the expected utility of player $i$ in the Nash equilibrium is at least $v_i - T > 0$, and thus $\lim_{x_i \rightarrow \underline{x}_{i}^+} u_i(x_i, F_{-i}) > 0$. 
        This means that $Supp(F_{-i})$ has a positive measure at $\underline{x}_{i}$.
        This indicates that both infimums of players' equilibrium strategies are the same and have positive measure. 
        However, for any $i$, player $i$ can bid $\underline{x}_{i} + \varepsilon$ so that $u_i(\underline{x}_{i} + \varepsilon, F_{-i}) > u_i(\underline{x}_{i}, F_{-i})$ for sufficiently small $\varepsilon$. This would be contradicted to the fact that player $i$'s support has a positive measure at $\underline{x}_i$.
        Therefore, we conclude that $T = L$.
        \item $B_2 < v_2$ and $B_1 \geq v_2$.
        Based on Case (2), it is easy to conclude that $L = \min\{B_1, B_2, v_1, v_2 \} = B_2$ is the supremum of the supports of players' strategies.
        \item $B_2 \geq v_2$ and $B_1 < v_2$.
        Based on Case (2), it is easy to conclude that $L = \min\{B_1, B_2, v_1, v_2 \} = B_1$ is the supremum of the supports of players' strategies.
    \end{enumerate}
    Therefore, we can conclude that $L = \min\{B_1, B_2, v_1, v_2 \} = \overline{x}_1 = \overline{x}_2$ in the Nash equilibrium.
\end{proof}

\vspace{5pt}
\noindent {\textbf{Proof of Lemma \ref{lem:support-single-battlefield-infimum-1}}}
\begin{proof}
    If $\underline{x}_i > 0$, then we have $\forall x_i \in [\underline{x}_i, \overline{x}_i)$, $u_i(x_i, \overline{x}_{-i}) < 0 = u_i(0, \overline{x}_{-i})$.
    This means that the utility obtained by player 1 when bidding 0 is higher than the utility obtained when bidding $x_i \in (0, \overline{x}_i)$.
    If player $i$ bids $x_i \in (\underline{x}_i, \overline{x}_i)$, then $u_i(x_i, \overline{x}_{-i}) < 0$.
    Therefore, we have $\forall x_i \in (\underline{x}_i, \overline{x}_i)$, $x_i \notin Supp(F_i)$.
    It implies that $Supp(F_i) = \{0, 1\}$.

    For player $-i$, we have
    \begin{align*}
        [F_i(0) + \frac{1}{2}(1 - F_i(0))] v_{-i} - L \geq \lim_{x_{-i} \rightarrow 0^+} u_{-i}(F_i, x_{-i}) = \lim_{\varepsilon \rightarrow 0^+} F_i(0) v_{-i} - \varepsilon,
    \end{align*}
    then we can get that $F_i(0) \leq 1 - \frac{2L}{v_{-i}}$ and $1 - F_i(0) \geq \frac{2L}{v_{-i}}$, which means that player $i$ bids $\underline{x}_i$ with a probability of no more than $1 - \frac{2L}{v_{-i}}$, and bids $\overline{x}_i$ with a probability of no less than $\frac{2L}{v_{-i}}$.
    It implies that $F_i(0) \leq 1 - \frac{2L}{v_{-i}}$.
    
    For player $i$, because of $0 \in Supp(F_i)$, then $i$'s utility is 0.
    For player $-i$, we can deduce that utility is at most $v_{-i} - 2L$.
\end{proof}

\vspace{5pt}
\noindent {\textbf{Proof of Lemma \ref{lem:support-single-battlefield-infimum-2}}}
\begin{proof}
    We first demonstrate that $\underline{x}_1 = \underline{x}_2$.
    We prove it by contradiction. 
    Without loss of generality, assume $\underline{x}_1 < \underline{x}_2$.
    For player 1, within the interval $[\underline{x}_1, \underline{x}_2)$, there must exist a bid $x$ in the support of player 1's strategy. 
    At this bid, we can derive that the utility of player 1 is $u_1(x, F_2) = -x$.
    Consider another bid $x' \in (\underline{x}_1, \underline{x}_2)$ and $x' < x$, if player 1 bids $x'$, we have that $u_1(x', F_2) = -x'$.
    Therefore, we have $u_1(x', F_2) > u_1(x, F_2)$, which means that if player 1's bid deviates from $x$ to $x'$, his utility increases, implying that $x$ is not in his strategy's support, which is a contradiction.
    
    We next prove that $0 = \underline{x}_1 = \underline{x}_2$.
    If $\underline{x}_1 = \underline{x}_2 > 0$, we consider three cases.
    \begin{itemize}
        \item Neither player has placed a mass point at $\underline{x}_1$.
        Then we have $\lim_{x_1 \rightarrow \underline{x}_i^+} u_1(x_1, F_2) = -\underline{x}_i < 0$, and $u_1(0, F_2) = 0$. 
        Obviously, player 1 can deviate from bidding $\underline{x}_1$ to 0 to achieve higher utility.
        \item Only one player, assuming player 1, places a mass at $\underline{x}_1$.
        Then player 1 gets negative utility at bid $\underline{x}_1$, and player 1 can benefit by deviating from bidding $\underline{x}_1$ to 0.
        \item Both players place masses at $\underline{x}_1$.
        Then we have that, for player 1, $u_1(\underline{x}_1, F_2) < \lim_{\varepsilon \rightarrow 0^+} u_1(\underline{x}_1 + \varepsilon, F_2)$, which means that player 1 can get more utility by transferring the mass at $\underline{x}_1$ to $\underline{x}_1 + \varepsilon$.
    \end{itemize}
    Therefore, there is always a player who increase utility by changing bid when $\underline{x}_1 = \underline{x}_2 > 0$.
    We can conclude that $\underline{x}_1 = \underline{x}_2 = 0$.
\end{proof}

\vspace{5pt}
\noindent {\textbf{Proof of Lemma \ref{lem:support-single-battlefield-continuously-nomass}}}
\begin{proof}
    Without loss of generality, let $B_1 \leq B_2$. According to our auction $\mathcal{G} = \{\{1, 2\}, \{1\}, B_1, B_2, v_{1}, v_{2}, u_1, u_2\}$, we can construct a new all-pay auction $\hat{\mathcal{G}} = \{\{1, 2\}, \{1\}, \hat{v}_{1}, \hat{v}_{2}, u_1, u_2\}$ with no budget constraint, where value $\hat{v}_1 = L$, $\hat{v}_2 = v_2$.
    
    Let $F_1, F_2$ be equilibrium strategies of player $1, 2$. 
    We can construct strategies of player 1 and 2,
    \begin{align*}
        \hat{F}_1 = F_1, \quad
        \hat{F}_2 = \begin{cases}
            \frac{F_2(x)}{F_2(L^{-})}, & x \in [0,L); \\
            1, & x = L.
        \end{cases}
    \end{align*}
    We have $u_1(x, \hat{F}_2) = \hat{F}_2(x) \cdot \hat{v}_1 - x = F_2(x) \cdot v_{i} - x = u_1(x, F_2)$.
    Obviously, when $\forall x \in Supp(\hat{F}_1)$ where $Supp(\hat{F}_1)$ is the support of strategy of player $1$, $u_1(x, \hat{F}_2)$ can be maximized.
    Similary, we have $u_2(\hat{F}_1, x) = \hat{F}_1(x) \cdot \hat{v}_2 - x = F_1(x) \cdot v_2 - x = u_1(x, F_2)$.
    When $\forall x \in Supp(\hat{F}_2)$ where $Supp(\hat{F}_2)$ is the support of strategy of player $2$, $u_2(\hat{F}_1, x)$ can be maximized.
    Thus for this new auction, by Lemma \ref{lem:support-single-battlefield-Baye-1996}, $\hat{F}_{1}$ and $\hat{F}_{2}$ are the equilibrium strategies of players 1 and 2, which are continuously randomized over the interval $(0, L)$ without any mass points. We can build a one-to-one mapping from set of bid in equilibrium strategies in $\mathcal{G}$ to those in $\hat{\mathcal{G}}$.

    Therefore, over the interval $(0, L)$, equilibrium strategy $F_1, F_2$ in our auction $\mathcal{G}$ and equilibrium strategy $\hat{F}_1, \hat{F}_2$ in the new auction $\hat{\mathcal{G}}$ share the same property of being continuous without any mass points.
\end{proof}

\vspace{5pt}
\noindent {\textbf{Proof of Propositon \ref{propositon-single-item}}}
\begin{proof}
    Let $B_1 = B_2 = 2$, and $v_1 = 4$, $v_2 = 3$, then $L = 2$.
    By possible structures of the support of the players' strategies, we consider three cases:
    \begin{itemize}
        \item Case 1: $Supp(F_1) = \{2\}$ and $Supp(F_2) = \{2\}$.
        Player 1's utility is 0, and player 2's utility is $-\frac{1}{2}$.
        Player 2 can change bid from 2 to 0, which increases utility from $-\frac{1}{2}$ to 0.
        Clearly, this cannot constitute a Nash equilibrium.
        \item Case 2: If $i = 1$, then $Supp(F_1) = \{0, L\}$ and $Supp(F_{2}) = \{L\}$.
        By Lemma \ref{lem:support-single-battlefield-infimum-1}, we can know that the probability of player 1 bidding 0 does not exceed $1 - \frac{2L}{v_{2}} < 0$.
        Obviously, it is impossible.
        If $i = 2$, then $Supp(F_2) = \{0, L\}$ and $Supp(F_{1}) = \{L\}$.
        By Lemma \ref{lem:support-single-battlefield-infimum-1}, we can know that the probability of player 2 bidding 0 does not exceed $1 - \frac{2L}{v_{1}} = 0$.
        When the probability of player 2 bidding 0 is exactly equal to 0, Case 2 reduces to Case 1.
        Therefore, Nash equilibrium does not exist.
        \item Case 3: $\overline{Supp(F_i)} = \{x | 0 \leq x \leq L\}$, $\forall i \in \{1, 2\}$.
        Note that both players do not place a mass at $L$.
        Because if player $i^* \in \{1, 2\}$ place a mass at $L$, then according to the tie-breaking rule, we have
        \begin{align*}
            \lim_{x \rightarrow L^-} u_{-i^*}(F_i, x) &= F_{i^*}(L^-)v_{-i^*} - L \\
            &< (F_{i^*}(L^-) + \frac{1}{2}(1 - F_{i^*}(L^-)))v_{-i^*} - L \\
            &= u_{-i^*}(F_i, L).
        \end{align*}
        It implies that $\exists \delta > 0$ such that $x \in (L - \delta, L)$, $u_{-i^*}(F_i, x) < u_{-i^*}(F_i, L)$.
        It means that the interval $(L - \delta, L) \cap Supp(F_{-i^*}) = \emptyset$, which contradicts Lemma  \ref{lem:support-single-battlefield-continuously-nomass}.
        Since both players do not place a mass at $L$, we can get $u_1 = 2$ and $u_2 = 1$.
        However, both players obtaining positive utility implies that both players have placed mass at 0.
        Hence, there will always be one player who moves the mass at 0 to $\epsilon > 0$ to increase utility as long as $\epsilon$ is small enough.
        This contradicts the fact that 0 is the infimum of supports of players' strategies.
        So, Nash equilibrium does not exist.
    \end{itemize}
    By analyzing all possible structure of support, we can conclude that there is no structure that can satisfy Nash equilibrium.
    Therefore, Nash equilibrium does not exist in this instance.
\end{proof}

\vspace{5pt}
\noindent {\textbf{Proof of Lemma \ref{lem:support-single-battlefield-endpoint-measure}}}
\begin{proof}
    For \textbf{Case 1}, we have $\min\{B_s, v_s\} > \min\{B_w, v_w\} = L$.
    In Nash equilibrium, player $s$'s utility is positive, which means that player $w$ places a mass at $0$.
    For player $s$, if he places a mass at 0, then when $\varepsilon$ is small enough, we have 
    \begin{align*}
        u_s(0, F_w) = \frac{F_w(0)}{2} v_s < F_w(0) v_s - \varepsilon = u_s(\varepsilon, F_w).
    \end{align*}
    It means that player $s$ can move the mass at 0 to $\varepsilon$, thereby increasing utility.
    So, player $s$ does not place a mass at $0$.
    Therefore, player $w$'s utility is $0$.
    We can get $\lim_{x \rightarrow L^-}u_w(F_s, x) = F_s(L^-) v_w - L = 0$, so $F_s(L^-) = \frac{L}{v_w}$.
    The probability that player $s$ bids at $L$ is $1 - \frac{L}{v_w}$.
    If player $w$ places a mass at $L$, then $\exists \delta > 0$ such that for $x \in (L - \delta, L)$, we have $F_w(x) < 1$, and
    \begin{align*}
        u_s(x, F_w) = F_w(x) v_s - (L - \delta) &= F_w(x) v_s - L + \delta \\
        &< v_s - L \\
        &= \lim_{\varepsilon \rightarrow 0^+} F_w(L + \varepsilon) v_s - (L + \varepsilon).
    \end{align*}
    Hence, we can derive that the utility for player $s$ when bidding $L + \varepsilon$ is higher than the utility when bidding $x \in (L - \delta, L)$.
    It implies that $x \in (L - \delta, L)$ is not best response to $F_w$, which contradicts with Lemma \ref{lem:support-single-battlefield-continuously-nomass}.
    Therefore, player $w$ does not place a mass at $L$.
    Then, we can get player $s$'s utility, that is $v_s - L$.
    The probability that player $w$ bids at $0$ can be calculated by the equation $v_s - L = F_w(0) v_s$, that is $\frac{v_s - L}{v_s}$.
    
    For \textbf{Case 2}, we consider two subcases:
    \begin{itemize}
        \item \text{subcase 2.1:} when $L = \min\{B_s, v_s\} < \min\{B_w, v_w\}$.
        In Nash equilibrium, player $w$'s utility is positive because player $w$ can ensure that his utility is at least $v_w - L - \varepsilon$ by bidding $L + \varepsilon$ as long as $\varepsilon$ is sufficiently small.
        So, we can deduce that player $s$ places a mass at 0, while player $w$ does not place a mass at 0.
        Then we konw that player $s$'s utility is 0.
        According to $0 = F_w(L^-)v_s - L$, we have $F_w(L^-) = \frac{L}{v_s} = 1$, implying player $w$ does not place a mass at $L$.
        According to the tie-breaking rule, we know that $s$ does not place a mass at $L$.
        Then, player $w$'s utility is $F_s(L) v_w - L = v_w - L$.
        Due to $v_w - L = F_s(0) v_w$, we have $F_s(0) = \frac{v_w - L}{v_w}$, implying that the probability that player $s$ bids at 0 is $\frac{v_{w} - L}{v_{w}}$.
        \item \text{subcase 2.2:} when $L = \min\{B_s, v_s\} = \min\{B_w, v_w\}$.
        We note that no player will place a mass at $L$.
        Hence, player $s$'s utility is 0, and player $w$'s utility is $v_w - L$.
        It means that the probability that player $s$ bids at 0 is $\frac{v_w - L}{v_w}$, and player $w$ does not place a mass at 0.
    \end{itemize}
\end{proof}

\vspace{5pt}
\noindent {\textbf{Proof of Lemma \ref{lem:support-single-battlefield-nash-equilibrium-support}}}
\begin{proof}
    For Case (1), we prove it by contradiction.
    Suppose there exists a player $i$ such that $\underline{x}_i < \overline{x}_i = L$.
    By Lemma \ref{lem:support-single-battlefield-infimum-1} and Lemma \ref{lem:support-single-battlefield-endpoint-measure}, there must be a player $i^*$ whose utility is 0.
    We consider two cases:
    \begin{itemize}
        \item If only one player has infimum less than supremum, then we have $Supp(F_{i^*}) = \{0, L\}$, $u_{i^*}(0, F_{-i^*}) = 0$ and $u_{i^*}(L, F_{-i^*}) = \frac{1}{2}v_{i^*} - L > 0$.
        This obviously contradicts $Supp(F_{i^*}) = \{0, L\}$.
        \item If the infimum of support of players' strategies is smaller than the supremum, then we have $Supp(F_{i^*}) = \{x| 0 \leq x \leq L\}$. We also have $u_{i^*}(L, F_{-i^*}) = (F_{-i^*}(L^-) + \frac{1}{2}(1 - F_{-i^*}(L^-)))v_{i^*} - L = (\frac{1}{2} + \frac{1}{2}F_{-i^*}(L^-))v_{i^*} - L > \frac{1}{2}v_{i^*} - L > 0$.
        This obviously contradicts $Supp(F_{i^*}) = \{x| 0 \leq x \leq L\}$.
    \end{itemize}
    Therefore, $\underline{x}_i = \overline{x}_i$ for any $i \in \{1, 2\}$.

    For Case (2), we prove it by contradiction.
    Suppose the infima of both players' strategy supports are 0.
    Then according to Lemma \ref{lem:support-single-battlefield-endpoint-measure}, there must exist a player $i^*$ whose utility is 0.
    Similar to the previous case, player $i^*$ can gain zero utility by bidding 0, while player $i^*$ can gain a positive utility by bidding $\overline{x}_{i^*} = L$, which contradicts our supposition.
    Therefore, at least one player must have an equal infimum and supremum for their strategy support.
    We now consider two cases:
    \begin{itemize}
        \item If every player has an equal infimum and supremum, then we have $Supp(F_{i'}) = \{L\}$ for any $i' \in \{1, 2\}$, and it is easy to verify that $(L, L)$ can form a Nash equilibrium by Lemma \ref{lem:pure-Nash-Equilibirum-single-item}.
        \item If only one player has an equal infimum and supremum, then we have $Supp(F_{i^*}) = \{0, L\}$, by Lemma \ref{lem:support-single-battlefield-infimum-1}, it is easy to verify that this can form a Nash equilibrium when player $i^*$ bids $L$ with a probability of no less than $\frac{2L}{v_{-i^*}}$ and 0 with a probability of no higher than $1 - \frac{2L}{v_{-i^*}}$, while player $-i^*$ bids $L$.
    \end{itemize}

    For Case (3), we prove it by contradiction.
    Suppose there exists a player whose support has an equal infimum and supremum. We consider two cases:
    \begin{itemize}
        \item if for some $i$, $\min\{B_i, v_i \} > \min\{B_{-i}, v_{-i} \}$, then we have $L \in Supp(F_i)$ and $L \in Supp(F_{-i})$.
        However, when player $-i$ bids $L$, its utility is negative, which contradicts $L \in Supp(F_{-i})$.
        \item if $\min\{B_1, v_1 \} = \min\{B_{2}, v_{2} \}$, then we consider three subcases:
        \begin{itemize}
            \item subcase 1: if $L = B_1$, then $B_1 = v_{2} = L$, $B_{2} > B_1$.
            Player 2's utility is $\frac{1}{2}v_{2} - L = \frac{1}{2} L - L < 0$ when he bids $L$.
            \item subcase 2: if $L = B_{2}$, then $B_{2} = v_1 = L$, $B_1 > L$.
            Player 1's utility is $\frac{1}{2}v_1 - L = \frac{1}{2} L - L < 0$ when he bids $L$.
            \item subcase 3: if $L \neq B_{1}$ and $L \neq B_{2}$, then $L = v_1 = v_{2}$.
            Player 1 and player 2 will both get negative utility when they bid $L$.
        \end{itemize}
        Therefore, the infimum of the support for both players' strategies is smaller than the supremum.
        Additionally, we can verify that if $Supp(F_i) = \{x|0 \leq x \leq L\}$ for any $i \in \{1, 2\}$, by Lemma \ref{lem:support-single-battlefield-endpoint-measure}, a Nash equilibrium can be formed.
    \end{itemize}
\end{proof}

\vspace{5pt}
\noindent {\textbf{Proof of Theorem \ref{thm:nash-equilibrium-single-battlefield}}}
\begin{proof}
    For \textbf{Case 1}, by Lemma \ref{lem:support-single-battlefield-nash-equilibrium-support}, it is easy to know that $\underline{x}_i = \overline{x}_i$ for any $i \in \{1, 2\}$.
    By Lemma \ref{lem:pure-Nash-Equilibirum-single-item}, we have that Nash equilibrium is $(B_1, B_2)$.
    
    For \textbf{Case 2}, by Lemma \ref{lem:support-single-battlefield-nash-equilibrium-support}, we have that
    \begin{itemize}
        \item $\underline{x}_i = \overline{x}_i$ for any $i \in \{1, 2\}$, by Lemma \ref{lem:pure-Nash-Equilibirum-single-item}, the Nash equilibrium is $(B_i, B_{-i})$.
        \item $\underline{x}_i = 0$ and $\underline{x}_{-i} = \overline{x}_{-i}$, by Lemma \ref{lem:support-single-battlefield-infimum-1}, the Nash equilibrium is that player $i$ bids 0 with a probability of no higher than $1 - \frac{2L}{v_{-i}}$ and bids $L$ with a probability of no less than $\frac{2L}{v_{-i}}$; player $-i$ bids $B_{-i}$.
    \end{itemize}

    For \textbf{Case 3}, by Lemma \ref{lem:support-single-battlefield-nash-equilibrium-support}, we have that $\underline{x}_i = 0$ for any $i \in \{1, 2\}$.
    By Lemma \ref{lem:support-single-battlefield-endpoint-measure}, we have that
    \begin{itemize}
        \item when $v_s > L$, for $x \in (0, L)$, $v_s - L = F_w(x)v_i - x$,  thence $F_w(x) = \frac{v_s - L + x}{v_i}$.
        For $x \in [0, L)$, $0 = F_s(x) v_w - x$, thence $F_s(x) = \frac{x}{v_w}$.
        Therefore, Nash equilibrium is as follows:
        \begin{align*}
            F_s(x) = \begin{cases}
                \frac{x}{v_w} \quad &x \in [0, L), \\
                1, & x = L,
            \end{cases}
            \quad
            F_w(x) = \begin{cases}
                \frac{v_s - L}{v_s}, & x = 0, \\
                \frac{v_s - L + x}{v_s}, & x \in (0, L].
            \end{cases}
        \end{align*}
        \item when $v_s = L$, for $x \in [0, L]$, $0 = F_w(x)v_s - x$, thence $F_w(x) = \frac{x}{v_s}$.
        For $x \in (0, L]$, $v_w - L = F_s(x) v_w - x$, thence $F_s(x) = \frac{v_w - L + x}{v_w}$.
        Therefore, Nash equilibrium is as follows:
        \begin{align*}
            F_s(x) = \begin{cases}
                \frac{v_w - L}{v_w}, & x = 0, \\
                \frac{v_w - L + x}{v_w}, & x \in (0, L],
            \end{cases}
            \quad
            F_w(x) = \frac{x}{v_s}, \quad x \in [0, L].
        \end{align*}
    \end{itemize}
\end{proof}

\vspace{10pt}
\noindent \textbf{B \quad Missing proofs in Section 4}

\vspace{5pt}
\noindent {\textbf{Proof of Theorem \ref{thm:two-item-nash-equilibrum-symmetric}}}
\begin{proof}
    We fix $f_1(x_{11}, x_{12})$, and the marginal distribution is $F_{1j}(x_{1j}) = \frac{x_{1j}}{c}$, $x_{1j} \in [0, c]$, $\forall j \in \{1, 2\}$.
    
    If player 2 bids $(x_{21}, x_{22})$ that satisfies $x_{21} \in [0, c]$ and $x_{22} \in [0, c]$, then player 2's utility is
    \begin{align*}
        F_{11}(x_{21}) v_{21} - x_{21} + F_{12}(x_{22}) v_{22} - x_{22} = (x_{21} + x_{22})(\frac{v_{21}}{c} - 1).
    \end{align*}
    Note that $0 \leq x_{21} + x_{22} \leq B_2$.
    If $c < v_{21}$, then we have $c = B_2$ and $\frac{v_{21}}{c} - 1 > 0$.
    So player 2 can bid $(x_{21}, x_{22})$ satisfying $x_{21} + x_{22} = B_2$ to maximize his utility.
    Obviously, the strategy $f_2(x_{21}, x_{22})$ we construct meets the condition where $x_{21} + x_{22} = B_2$.
    If $c = v_{21}$, then we have $\frac{v_{21}}{c} - 1 = 0$.
    So regardless of how player 2 bids, player 2's utility is always 0.
    Obviously, the strategy $f_2(x_{21}, x_{22})$ we constructed is a best response to $f_1$ when $x_{21} \in [0, c]$ and $x_{22} \in [0, c]$.
    
    If player 2 bids $(x_{21}, x_{22})$ that satisfies $x_{2j} > c$, $\exists j \in \{1, 2\}$.
    We have two cases:
    \begin{itemize}
        \item Case 1: $c = B_2$. It is impossible to bid $x_{2j} > c$.
        \item Case 2: $c = v_{2j}$. Then on the item 2, player 2's utility is $v_{2j} - x_{2j} < 0$.
    \end{itemize}
    Hence, we can deduce that if $x_{2j} > c$, $\exists j \in \{1, 2\}$, player 2's utility is less than utility of the case where $x_{2j} \in [0, c]$, for any $j \in \{1, 2\}$.
    Therefore, we can conclude that $f_2$ is a best response to $f_1$.
    
    Similarly, fixing $f_2$, we can conclude that $f_1$ is a best response to $f_2$.
    Therefore, the strategy profile $(f_1, f_2)$ is a Nash equilibrium.
\end{proof}

\vspace{5pt}
\noindent {\textbf{Proof of Theorem \ref{thm:nash-equilibrium-two-items-case1}}}
\begin{proof}
    Before proving it, we need to note the following facts: $x_{i1} + x_{i2} \leq B_i$, $\forall i \in \{s, w\}$, $L_1 = \min\{v_{w1}, B_w\}$ and $L_2 = \min\{v_{w2}, B_w\}$.
    
    First, we prove $f_w(x_{w1}, x_{w2})$ is a best response to $f_s(x_{s1}, x_{s2})$.
    
    We calculate the marginal distribution of player $s$ over the two items, which are:
    \begin{align*}
        F_{s1}(x_{s1}) = \begin{cases}
            \frac{x_{s1}}{v_{w1}}, \quad &x_{s1} \in [0, L_1); \\
            1, \quad &x_{s1} = L_1;
        \end{cases}
        \quad
        F_{s2}(x_{s2}) = \begin{cases}
            \frac{x_{s2}}{v_{w2}}, &x_{s2} \in [0, L_2); \\
            1, &x_{s2} = L_2.
        \end{cases}
    \end{align*}
    When player $w$ bids $(x_{w1}, x_{w2})$ satisfying $x_{w1} \in [0, L_1]$ and $x_{w2} \in [0, L_2]$, the utility is as follows:
    \begin{align*}
        u_w(f_s(x_{s1}, x_{s2}), (x_{w1}, x_{w2})) &= u_{w1}(F_{s1}(x_{s1}), x_{w1}) + u_{w2}(F_{s2}(x_{s2}), x_{w2}) \\
        &= \frac{x_{w1}}{v_{w1}} v_{w1} - x_{w1} + \frac{x_{w2}}{v_{w2}} v_{w2} - x_{w2} \\
        &= 0.
    \end{align*}
    It implies that as long as player $w$'s bid $x_{wj}$ satisfies $x_{wj} \leq L_j$, their utility on item $j$ is always 0, regardless of the value of $L_j$.
    It is clear that the strategy $f_w(x_{w1}, x_{w2})$ we constructed satisfies $x_{wj} \leq L_j$.
    Therefore, $f_w(x_{w1}, x_{w2})$ is a best response to $f_s(x_{s1}, x_{s2})$ when $x_{w1} \in [0, L_1]$ and $x_{w2} \in [0, L_2]$.
    
    When the situation occurs where $x_{wj} > L_j$, $\exists j \in \{1, 2\}$, according to the different values of $L_j$, we further divide it into the following four situations, and we analyze each situation in detail.
    \begin{itemize}
        \item Case (1.1): $L_1 = B_w$ and $L_2 = B_w$;
        \item Case (1.2): $L_1 = B_w$ and $L_2 = v_{w2}$;
        \item Case (1.3): $L_1 = v_{w1}$ and $L_2 = v_{w2}$;
        \item Case (1.4): $L_1 = v_{w1}$ and $L_2 = B_w$.
    \end{itemize}
    Dut to $x_{w1} + x_{w2} \leq B_s$, the Case (1.1) will not happen.
    
    For Case (1.2), $L_1 = B_w$ and $L_2 = v_{w2}$, we have the following subcases:
    \begin{itemize}
        \item subcases 1.2.1: $x_{w1} > L_1 = B_w$ and $x_{w2} > L_2$. Obviously, it can not happen.
        \item subcases 1.2.2: $x_{w1} \leq L_1 = B_w$ and $x_{w2} > L_2$. Obviously, player $w$'s utility is 0 on the item 1 and negative on the item 2.
        The total utility is negative.
        \item subcases 1.2.3: $x_{w1} > L_1 = B_w$ and $x_{w2} \leq L_2$. Obviously, it can not happen.
    \end{itemize}

    For Case (1.3), $L_1 = v_{w1}$ and $L_2 = v_{w2}$, we have the following subcases:
    \begin{itemize}
        \item subcases 1.3.1: $x_{w1} > L_1$ and $x_{w2} > L_2$. Obviously, player $w$'s utility is negative on the item 1 and item 2.
        \item subcases 1.3.2: $x_{w1} \leq L_1$ and $x_{w2} > L_2$. Obviously, player $w$'s utility is 0 on the item 1 and negative on the item 2.
        The total utility is negative.
        \item subcases 1.3.3: $x_{w1} > L_1$ and $x_{w2} \leq L_2$. Obviously, player $w$'s utility is negative on the item 1 and 0 on the item 2.
        The total utility is negative.
    \end{itemize}

    For Case (1.4), $L_1 = v_{w1}$ and $L_2 = B_w$, this is similar to Case (2).

    Therefore, if $\exists j \in \{1, 2\}$, player $w$ bids $x_{wj} > L_j$, player $w$ always gets negative utility.
    Player $w$ will not bid more than $L_j$ on any item $j$.
    Hence, $f_w(x_{w1}, x_{w2})$ is a best response to $f_s(x_{s1}, x_{s2})$ for any $(x_{w1}, x_{w2})$ that satisfies $x_{w1} + x_{w2} \leq  B_w$.
    
    Next, we prove $f_s(x_{s1}, x_{s2})$ is a best response to $f_w(x_{w1}, x_{w2})$.
    We calculate the marginal distribution of player $w$ over the two items, which are:
    \begin{align*}
        F_{w1}(x_{w1}) = \begin{cases}
            1 - \frac{L_1}{v_{s1}}, &x_{w1} = 0; \\
            1 - \frac{L_1}{v_{s1}} + \frac{x_{w1}}{v_{s1}}, &x_{w1} \in (0, L_1];
        \end{cases}
    \end{align*}
    \begin{align*}
        F_{w2}(x_{w2}) = \begin{cases}
            1 - \frac{L_2}{v_{s2}}, &x_{w2} = 0; \\
            1 - \frac{L_2}{v_{s2}} + \frac{x_{w2}}{v_{s2}}, &x_{w2} \in (0, L_2].
        \end{cases}
    \end{align*}
    When player $s$ bids $(x_{s1}, x_{s2})$ satisfying $x_{s1} \in [0, L_1]$ and $x_{s2} \in [0, L_2]$, the utility is as follows:
    \begin{align*}
        u_s((x_{s1}, x_{s2}), f_w(x_{w1}, x_{w2})) &= u_{s1}(x_{s1}, F_{w1}(x_{w1})) + u_{s2}(x_{s2}, F_{w2}(x_{w2})) \\
        &= (1 - \frac{L_1}{v_{s1}} + \frac{x_{s1}}{v_{s1}}) v_{s1} - x_{s1} + (1 - \frac{L_2}{v_{s2}} + \frac{x_{s2}}{v_{s2}}) v_{s2} - x_{s2} \\
        &= v_{s1} - L_1 + v_{s2} - L_2.
    \end{align*}
    It implies that as long as player $s$'s bid $x_{sj}$ satisfies $x_{sj} \leq L_j$, their utility on item $j$ is always $v_{s1} - L_1 + v_{s2} - L_2$.
    It is clear that the strategy $f_s(x_{s1}, x_{s2})$ we constructed satisfies $x_{sj} \leq L_j$.
    Therefore, $f_s(x_{s1}, x_{s2})$ is a best response to $f_w(x_{w1}, x_{w2})$ when $x_{s1} \in [0, L_1]$ and $x_{s2} \in [0, L_2]$.
    If $\exists j \in \{1, 2\}$ such that $x_{sj} > L_j$, then player $s$'s utility on the item $j$ is $v_{sj} - x_{sj}$, which is less than $v_{sj} - L_j$.
    It means that on item $j$, as long as player $s$'s bid exceeds $L_j$, his utility will be less than $v_{sj} - L_j$, which will cause his total utility to be less than $v_{s1} - L_1 + v_{s2} - L_2$.
    Therefore, we can conclude that on every item, player $s$'s bid does not exceeds $L_j$.
    Hence, $f_s(x_{s1}, x_{s2})$ is a best response to $f_w(x_{w1}, x_{w2})$ for any $(x_{s1}, x_{s2})$ that satisfies $x_{s1} + x_{s2} \leq  B_s$.
\end{proof}

\begin{theorem}\label{thm:nash-equilibrium-two-items-case4}
    Given the profile $(B_1, B_2, v_{11}, v_{12}, v_{21}, v_{22})$ with $B_1 \neq B_2$, if $v_{s1} = L_1$ and $v_{s2} > L_2$, then the following strategies form a Nash equilibrium:
    \begin{align*}
        f_s(x_{s1}, x_{s2}) = \begin{cases}
            \frac{1}{v_{w1}}, &x_{s1} = (0, T_1], \, x_{s2} = L_2; \\
            \frac{\frac{L_2}{v_{w2}}}{\sqrt{L_2^2 + (L_1 - T_1)^2}}, & \begin{array}{l}
                x_{s1} \in (T_1, L_1], \, x_{s2} \in [0, L_2), \\
                x_{s2} = \frac{L_2}{T_1 - L_1} x_{s1} - \frac{L_2}{T_1 - L_1} L_1,
            \end{array}
        \end{cases}
    \end{align*}
    where $T_1 = -\frac{L_2}{v_{w2}} v_{w1} + L_1$.
    When $x_{s1} = 0, \, x_{s2} = L_2$, there is a mass point, the probability value is $1 - \frac{L_1}{v_{w1}}$.
    \begin{align*}
        f_w(x_{w1}, x_{w2}) = \begin{cases}
            \frac{1}{v_{s1}}, &x_{w1} \in [T_3, L_1], \, x_{w2} = 0; \\
            \frac{\frac{L_2}{v_{s2}}}{\sqrt{L_2^2 + T_3^2}}, &x_{w1} \in [0, T_3), \, x_{w2} \in (0, L_2], x_{w2} = - \frac{L_2}{T_3} x_{w1} + L_2,
        \end{cases}
    \end{align*}
    where $T_3 = L_1 - v_{s1} + \frac{L_2}{v_{s2}} v_{s1}$.
\end{theorem}

\vspace{10pt}
\noindent \textbf{\textbf{C \quad Missing proofs in Section 4}}

\vspace{5pt}
\noindent {\textbf{Proof of Theorem \ref{thm:nash-equilibrium-three-items}}}
\begin{proof}
    Let $L_j = v_j$ for $j \in \{1, 2, 3\}$.
    For Case 1, we examine the strategies of the two players, and each is the best response to the other.
    Fix player 1's strategy $f_1$, we can calculate the marginal distribution, that is $\forall j \in \{1, 2, 3\}$, $F_{1j} = \frac{x}{L_j}$ for $x \in [0, L_j]$.
    For player 2's bid $(x_{21}, x_{22}, x_{23})$ satisfying $x_{21} + x_{22} + x_{23} \leq B_2$, on item $j$, if $x_{2j} \in [0, L_j]$, then the utility from item $j$ is $u_{2j}(F_{1j}, x_{2j}) = F_{1j}(x_{2j}) v_j - x_{2j} = \frac{x_{2j}}{L_j} v_{j} - x_{2j} = 0$.
    Therefore, player 2's utility is always 0 as long as $x_{2j} \in [0, L_j]$.
    Clearly, in the strategy $f_2$ we have constructed, for every item $j$, $x_{2j} \in [0, L_j]$ holds.
    However, for item $j$, if $x_{2j} > L_j$, then we have $u_{2j}(F_{1j}, x_{2j}) = v_j - x_{2j} < 0$.
    This means that on item $j$, the bid of player 2 will not exceed $L_j$.
    We can conclude that $f_2$ we have constructed is a best response to $f_1$.
    Similarly, fix player 2's strategy $f_2$, we can also prove that $f_1$ is a best response to $f_2$.
    Hence, the strategy profile $(f_1, f_2)$ is a Nash equilibrium.

    For Case 2, the proof process is similar to that of Case 1 and is therefore omitted.
\end{proof}

\end{document}